%% file: main.tex
\documentclass[sigconf,nonacm]{acmart}

\usepackage{graphicx} % Required for inserting images
\usepackage{subcaption}
\usepackage{multirow}
\usepackage{pifont}

\newcommand{\NAME}{\textsf{gLLM}}
\newcommand{\METHODT}{\textsf{Token Throttling}}
\newcommand{\CHUNKED}{\textsf{Sarathi-Serve}}
\newcommand{\VLLM}{\textsf{vLLM}}
\newcommand{\SGLANG}{\textsf{SGLang}}

\newcommand{\xwz}[2]{\textcolor{black}{#1}} % blue
\newcommand{\gty}[2]{\textcolor{black}{#1}} % olive

\author{Tianyu Guo}
\email{guoty9@mail2.sysu.edu.cn}
\orcid{0009-0005-2979-4486}
\affiliation{
    \institution{Sun Yat-sen University} 
    \city{Guangzhou}
    \country{China}
}

\author{Xianwei Zhang\textsuperscript{\#}}
\email{zhangxw79@mail.sysu.edu.cn}
\orcid{0000-0003-3507-4299}
\affiliation{
    \institution{Sun Yat-sen University} 
    \city{Guangzhou}
    \country{China}
}

\author{Jiangsu Du}
\email{dujiangsu@mail.sysu.edu.cn}
\orcid{0000-0003-4707-9492}
\affiliation{
    \institution{Sun Yat-sen University} 
    \city{Guangzhou}
    \country{China}
}

\author{Zhiguang Chen}
\email{chenzhg29@mail.sysu.edu.cn}
\orcid{0000-0002-9318-5715}
\affiliation{
    \institution{Sun Yat-sen University} 
    \city{Guangzhou}
    \country{China}
}

\author{Nong Xiao}
\email{xiaon6@mail.sysu.edu.cn}
\orcid{0000-0002-2166-977X}
\affiliation{
    \institution{Sun Yat-sen University} 
    \city{Guangzhou}
    \country{China}
}

\author{Yutong Lu}
\email{luyutong@mail.sysu.edu.cn}
\orcid{0000-0001-5315-3375}
\affiliation{
    \institution{Sun Yat-sen University} 
    \city{Guangzhou}
    \country{China}
}

\begin{document}

\title{\NAME: \underline{G}lobal Ba\underline{l}anced Pipe\underline{l}ine Parallelis\underline{m} System for Distributed LLM Serving with \METHODT}

% Pipeline parallelism has been a predominant method for deploying large language models (LLMs) across distributed nodes. Although it features with high throughput in serving requests, its performance is often limited by pipeline bubbles. Towards the issue, existing methods like Sarathi-Serve have been proposed by hybrid scheduling chunked prefill and decode tokens. However, the number of tokens scheduled each time may experience significant fluctuations. To address the inefficiency, we propose gLLM, a global balanced systems with Tokens Throttling to mitigate the pipeline bubbles. Tokens Throttling is a fine-grained schedule policy that regulates the number of prefill and decode tokens separately, allowing for balanced computation based on global information from the inference systems. Moreover, gLLM adopts an asynchronous execution and message passing architecture. Experiments on representative LLMs show that gLLM can improve the maximum throughput by 113%∼398% compared to the state of the art systems with lower latency.

\begin{abstract}
Pipeline parallelism has \xwz{emerged as}{been} a predominant \xwz{approach}{method} for deploying large language models (LLMs) across distributed nodes, \xwz{owing}{primarily due} to its lower communication overhead compared to tensor parallelism. 
\xwz{While}{Although} \xwz{demonstrating}{it features with} high throughput in \xwz{request serving}{serving requests}, \xwz{}{its}\xwz{pipeline parallelism often suffers from}{} performance \xwz{limitations caused}{is often limited} by pipeline bubbles, which are \xwz{primarily resulted from}{typically caused by} imbalanced computation delays \xwz{across batches}{between each batch}. % \xwz{that degrades}{, degrading} overall efficiency.
\xwz{}{Towards the issue,} Existing methods like \CHUNKED~\xwz{attempt to address this through}{have been proposed by} hybrid scheduling \xwz{of}{} chunked prefill and decode tokens \xwz{using}{with} a fixed token budget. 
However, \xwz{such methods}{the number of tokens scheduled each time} may experience significant fluctuations due to \xwz{either}{} insufficient prefill tokens or uneven distribution of decode tokens, \xwz{ultimately}{} leading to computation\xwz{al}{} imbalance.
To \xwz{overcome these inefficiencies}{address the inefficiency}, we \xwz{present}{propose} \NAME, a global\xwz{ly}{} balanced \gty{pipeline parallelism}{} system\xwz{}{s} \xwz{incorporating}{with} \METHODT~to \xwz{effectively}{} mitigate the pipeline bubbles. \xwz{Our}{} \METHODT~\xwz{mechanism}{} is a fine-grained schedul\xwz{ing}{e} policy that \xwz{independently}{} regulates the \xwz{quantities}{number} of prefill and decode tokens\xwz{}{ separately}, \xwz{thus enabling}{allowing for} balanced computation \xwz{by leveraging}{based on} global information from the inference system\xwz{}{s}. 
Specifically, for decode tokens, \xwz{\NAME~maintains near-consistent token count across processing batches}{it ensures the batched token count reamins nearly consistent across batches}. 
For prefill tokens, it dynamically adjusts \xwz{batch sizes based on both}{the batch size according to the} total pending tokens and the memory utilization rate\xwz{s}{} of \xwz{}{the} key-value cache (KV cache).
\xwz{Furthermore}{Moreover}, \NAME~\gty{runtime}{} adopts an asynchronous execution and message passing architecture \xwz{specifically optimized for}{according to the characteristics of} pipeline parallelism \xwz{characteristics}{}.
\xwz{Experimental evaluations with}{Experiments on} representative LLMs show that \NAME~\xwz{achieves significant performance improvements, delivering 11\% to 398\% higher maximum throughput compared to state-of-the-art pipeline or tensor parallelism systems, while simultaneously maintaining lower latency}{can improve the maximum throughput by 11\%$\sim$398\% compared to the state of the art pipeline parallelism systems or tensor parallelism systems with low latency}. We have open-sourced our code at \url{https://github.com/gty111/gLLM}.
\end{abstract}

\maketitle

\begingroup\renewcommand\thefootnote{\#}
\footnotetext{Corresponding author.}
\endgroup

\input{chapters/1.introduction}
\input{chapters/2.background_motivation.tex}

\input{chapters/4.design}
\input{chapters/6.result_analysis}
\input{chapters/7.related_work}

\input{chapters/8.conclusion}

\bibliographystyle{ieeetr}
\bibliography{ref}

\end{document}

%% file: chapters/1.introduction.tex
\section{INTRODUCTION}
Large language models (LLMs) have demonstrated \xwz{remarkable capabilities in performing}{the ability to perform} complex tasks like logical reasoning \cite{DBLP:conf/acl/SunXL0WSWY24,DBLP:conf/emnlp/0001H24a,DBLP:conf/emnlp/ToroghiGPS24a}, \xwz{mathematical}{math} problem\xwz{}{s} solving \cite{DBLP:conf/acl/0001LLKNC0T0C24,DBLP:conf/acl/LiCZKB24,DBLP:conf/acl/LiuZQDFZZZLC24,DBLP:conf/acl/LuZRWSPZL24,DBLP:conf/nips/DidolkarGKGVLRB24,DBLP:conf/nips/SetlurGGGSK24} and agent acting \cite{DBLP:conf/aaai/SchumannZFFRW24,DBLP:conf/aaaiss/0001PWM24,DBLP:conf/aaaiss/ToukmajiT24,DBLP:conf/acl/DengXSLT0L000S24,DBLP:conf/acl/ShiSYCL24,DBLP:conf/acl/YangWCWPGHS024}. 
As \xwz{model parameters scale to}{the number of model parameter reaches} hundreds of billion or even trillions \cite{DBLP:journals/corr/abs-2412-19437,DBLP:journals/corr/abs-2407-21783,DBLP:journals/corr/abs-2311-16867,DBLP:journals/corr/abs-2303-08774,DBLP:journals/corr/abs-2403-05530,DBLP:journals/jmlr/FedusZS22}, distributed serving of LLMs 
\cite{DBLP:journals/corr/abs-2408-00741,DBLP:journals/corr/abs-2406-01566,DBLP:conf/nips/BorzunovRCBDBSR23,DBLP:journals/corr/abs-2312-04025} \xwz{has become essential}{becomes necessary} due to\xwz{}{the limited} GPU memory \xwz{constraints}{}. 
\xwz{Among distributed approaches,}{} pipeline parallelism has emerged as a \xwz{pre}{}dominant \xwz{method}{approach} for\xwz{}{distributed} training \xwz{and}{or} serving models like deep nerual networks (DNN) \cite{DBLP:journals/corr/abs-1806-03377,DBLP:conf/icml/NarayananPSCZ21,DBLP:conf/nips/HuangCBFCCLNLWC19} \xwz{and}{or} LLMs \cite{DBLP:conf/asplos/SunCWF0WC24,DBLP:conf/sc/LiuCZ023,DBLP:conf/ppopp/LinL0WZZ25,DBLP:conf/ppopp/LiuLTJ25,DBLP:conf/iclr/QiWHL24,DBLP:conf/icml/KimKYC23,DBLP:conf/sc/NarayananSCLPKV21,DBLP:conf/osdi/AgrawalKPMKGTR24,DBLP:conf/isca/PatelCZSGMB24,DBLP:conf/osdi/ZhongLCHZL0024}, \xwz{primarily due}{thanks} to its low communication overhead \xwz{that}{, which} effectively \xwz{mitigates}{addresses} bandwidth limitations.
However, \xwz{this method}{it} often incurs pipeline bubbles, \xwz{i.e.,}{} periods of GPU inactivity \xwz{where}{, because} subsequent pipeline stages \xwz{must}{have to} wait for prior ones to complete \xwz{micro-batch processing}{processing of their respective micro-batches}. 
\xwz{These inefficient bubbles mainly stem from computational imbalance between micro-batches}{
Pipeline bubbles often arised by unbalanced computation between each micro-batch}. 
While existing \xwz{researches have focused on}{approaches targeting LLMs are mostly for} optimizing pipeline bubbles in training \xwz{scenarios}{} \cite{DBLP:conf/asplos/SunCWF0WC24,DBLP:conf/sc/LiuCZ023,DBLP:conf/ppopp/LinL0WZZ25,DBLP:conf/ppopp/LiuLTJ25,DBLP:conf/iclr/QiWHL24,DBLP:conf/icml/KimKYC23,DBLP:conf/sc/NarayananSCLPKV21,DBLP:conf/osdi/AgrawalKPMKGTR24}, \xwz{recent}{other} methods like \CHUNKED~\cite{DBLP:conf/osdi/AgrawalKPMKGTR24} and prefill-decode disaggregated architectures \cite{DBLP:conf/isca/PatelCZSGMB24,DBLP:conf/osdi/ZhongLCHZL0024} \xwz{aim}{try} to \xwz{address computational imbalances specifically between prefill and decode stages}{solve the problems raised by unbalanced computation in prefill and decode stage}. \xwz{Our analysis reveals that these solutions remain insufficient for effectively resolving the bubble issues}{ We illustrate that they both fail to solve the problems effectively}.

\begin{figure}
    \centering
    \includegraphics[width=\linewidth]{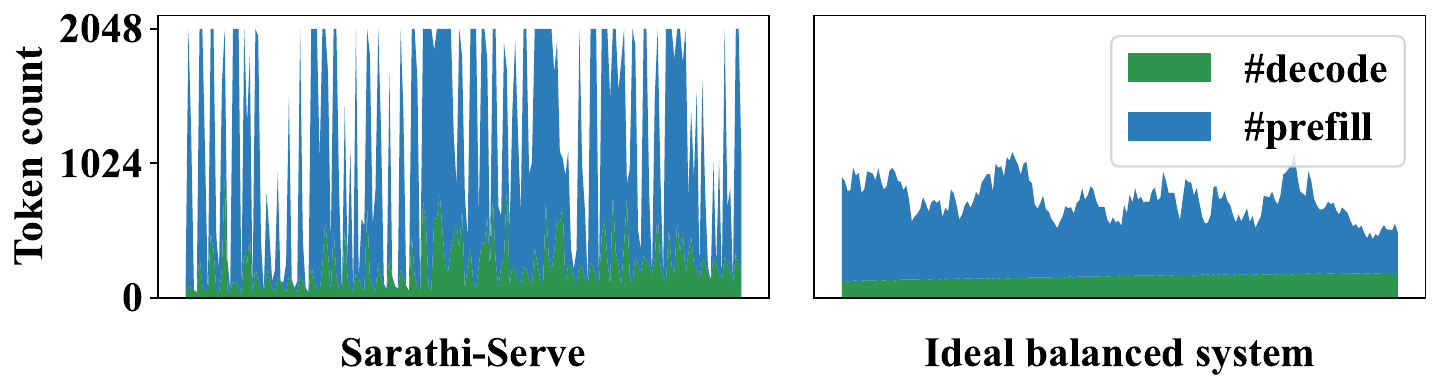}
    \caption{A comparison of the scheduled token counts of prefill and decode stage between \CHUNKED~and an ideal balanced system per iteration (horizontal axis), with the maximum batched token count (token budget) set to 2048 for both systems.}
    \label{fig:schedule_tokens}
\end{figure}

Each LLM serving request \xwz{undergoes two distinct stages}{experience two stage computation}. 
The first \xwz{stage,}{stage is} prefill, \xwz{processes the prompt tokens to compute keys and values (KV cache), then generating the initial}{ which computes the keys and values (KV cache) of the prompt tokens and generate first} output token. 
\xwz{The second stage,}{And the second stage is} decode, \xwz{produces the remaining}{which generates the rest of} output tokens. 
\xwz{The two stages exhibit markedly different computational characteristics, where prefill operations typically saturate GPU capacity, while decode operations demonstrate significantly lower compute utilization}{
Prefill computation tends to saturate the GPU. 
Instead, decode computation has low compute utilization}. 
Therefore, \xwz{to enhance hardware efficiency, multiple decode tokens are commonly batched together for parallel processing}{many decode tokens are batched together to compute the next tokens}. 
\xwz{However}{Nevertheless}, the interleaved execution of prefill and decode \xwz{operations}{} may \xwz{create mutual interference}{interfere with each other}. 
To mitigate the unbalanced batching policy, \CHUNKED~\cite{DBLP:conf/osdi/AgrawalKPMKGTR24} \gty{proposes}{is proposed} \xwz{to apply a}{for allowing} hybrid \xwz{scheduling that allocates a fixed token budget between chunked prefill and decode operations}{schedule chunked prefill and decode tokens with a fixed token budget}. 
However, \xwz{this solution fails to properly account for}{it overlooks} the \xwz{critical}{} prefill-to-decode ratio. \xwz{In practice}{and in real-world scenarios}, decode tokens \xwz{often}{frequently} lack sufficient \xwz{corresponding prefill tokens for effective co-scheduling}{prefill tokens for scheduling together}.
Figure \ref{fig:schedule_tokens} \xwz{illustrates the disparity in}{presents the breakdown of} batched token counts per iteration between \CHUNKED~and an \xwz{optimally}{ideal} balanced system. 
\xwz{It can be observed}{We observe} that \CHUNKED's \xwz{scheduling results in substantially greater token count volatility compared to the balanced counterpart}{scheduled token counts exhibit significantly higher fluctuations than the balanced system}. 
These fluctuations \xwz{can be attributed}{occur due} to two \xwz{primary}{} factors: (1) \xwz{missed opportunities for batching decode tokens with prefill tokens, and}{decode tokens lack the opportunity to be batched with prefill tokens} (2) \xwz{uneven}{imbalanced} distribution of decode tokens across batches.
Unfortunately, they frequently \xwz{induce}{create} pipeline bubbles that \xwz{significantly}{} degrade system performance. 
\xwz{Whereas}{Although} reducing\xwz{}{the} token budget could \xwz{theoretically smooth}{mitigate} these fluctuations, \xwz{such approach}{it} would \xwz{disproportionately penalize}{also lower the} prefill rates, ultimately \xwz{constraining}{limiting} the \xwz{overall system}{system's overall} throughput.

\xwz{To address the divergent computational demands between prefill and decode stages}{To confront the distinct computation characteristic between prefill and decode stage}, prefill-decode disaggregated architectures
\gty{}{recent work has proposed prefill-decode disaggregated architectures}
\cite{DBLP:conf/isca/PatelCZSGMB24,DBLP:conf/osdi/ZhongLCHZL0024} 
\xwz{have been recently}{are also} proposed to \xwz{alleviate}{eliminate} prefill-decode interference and achieve low cost budget for each stage respectively. 
\xwz{Generally, the designs}{They} assign prefill and decode computation to different nodes \xwz{connected via}{and connect the computation of two stage by transmitting} KV cache \xwz{transmission}{}. 
Despite separate \gty{computation}{scheduling} of the prefill and decode stages, computational imbalance persists \gty{across either}{either across} prefill batches or decode batches.
\gty{}{Additionally, they introduce non-negligible KV cache transmission overhead, which is proportional to the length of the sequence. 
This becomes a great concern for deploying LLMs in clusters with low communication bandwidth between nodes, which incurs high latency for transmitting KV cache of long context \cite{DBLP:conf/sosp/WuLZ0L024,DBLP:conf/icml/TangZZXKH24}.} 
Moreover, determining the optimal ratio of GPUs allocated to the prefill stage versus the decode stage becomes challenging under dynamically fluctuating request rates. 
The tight coupling between prefill and decode nodes also raises fault tolerance concerns. 
A failure in one stage could cascade to the other. 
%However, this approach still suffers from computational imbalance across batches, significant KV cache transmission overhead that scales with sequence length (particularly problematic for long-context LLMs in low-bandwidth clusters [39,40]), and challenges in dynamically allocating GPU resources between stages under fluctuating workloads, compounded by systemic fragility where node failures in one stage can cascade to the other. These limitations collectively hinder the architecture's ability to deliver both computational efficiency and operational robustness for production-scale LLM deployment.

To effectively \gty{address}{solve} the unbalanced problems and \gty{minimize}{further reduce} pipeline bubbles, we design \NAME, a \gty{globally}{global} efficient \gty{pipeline parallelism}{} system that balances computation across micro-batches \gty{using}{through} \METHODT. 
This mechanism intelligently regulates the number of \gty{}{batched} tokens \gty{processed in the}{} prefill and decode stage\gty{}{separately} based on \gty{real-time}{global} system states. 
For decode tokens, it \gty{distributes tokens evenly across micro-batches}{considers the total count of tokens under decoding}. 
For prefill tokens, \gty{it balances batch tokens based on}{it accounts for} both the pending prefill tokens and KV cache utilization rate\gty{s}{}. 
\gty{By dynamically throttling tokens count}{In that case}, \gty{\NAME~}{it}\gty{achieves}{enables better hybrid execution of prefill and decode operations, achieving} better load balancing across micro-batches \gty{and greatly alleviates the pipeline bubble issue}{}.
\gty{Moreover}{Besides}, considering the characteristic of pipeline parallelism, \gty{\NAME~runtime}{the implementation of \NAME} adopts an asynchronous execution and message passing architecture for reducing data dependency \gty{and CPU overhead}{}.

The contributions of this paper are:
\begin{itemize}
    \item We highlight the observations that pipeline bubbles caused by unbalanced computation in prefill and decode stage significantly degrade the performance of inference systems.
    \item We present \NAME, a distributed serving system with \METHODT~for effectively balancing the computation across batches to reduce pipeline bubbles. 
    \item \METHODT~dynamically adjusts the batch size of the prefill and decode stage separately to achieve balanced schedule based on real-time inference system feedback.
    \item Experiments on representative LLMs show that \NAME~enhances maximum throughput by 11\% to 398\% compared \gty{to}{with} \gty{state-of-the-art}{the state of the art} pipeline\gty{}{parallelism systems} or tensor parallelism systems, while simultaneously achieving lower latency.
\end{itemize}

%% file: chapters/2.background_motivation.tex
\section{BACKGROUND AND MOTIVATION}

\subsection{The Transformer Architecture}

The predominant architecture in contemporary LLMs adopts the decoder-only transformer \cite{DBLP:conf/nips/VaswaniSPUJGKP17}. 
In which, the input begins with an embedding layer, which converts token IDs into hidden states while incorporating positional encoding to keep sequence order. 
These hidden states then pass through multiple decoder layers, each consisting of a self-attention mechanism (with causal masking to ensure autoregressive properties), layer normalization, and a multi-layer perceptron (MLP). 
Between them, self-attention is the core of the transformer, enabling LLMs to capture long-range contextual dependencies. 
After processing through the decoder layers, a linear projection layer transforms the final hidden states into logits, representing the probability distribution over the vocabulary. 
Finally, a sampling strategy (e.g., greedy, top-k, or nucleus sampling) selects the next token for generation. 
% Although decoder layers account for the majority of computational overhead, auxiliary operations like logits projection and token sampling contribute non-negligible latency during inference.

\subsection{LLM Inference Procedure}

\textbf{Autoregressive Decoding.} The procedure of LLM inference is autoregressive where each token is generated based on all preceding \gty{onces}{tokens} through the computation of attention scores between their keys and values.
To avoid the recomputation of keys and values, they are retained for subsequent steps \cite{DBLP:conf/osdi/YuJKKC22,DBLP:conf/sosp/KwonLZ0ZY0ZS23}. 
Base on that, LLM inference can be divided into two distinct phases: 
1. Prefill: The prompt's tokens are processed in parallel to populate the KV cache and generate the first output token. 
This phase fully utilizes GPU compute resources due to high parallelism. 
2. Decode: Each new token is generated sequentially by attending to the last token and the stored KV cache. 
This phase often underutilizes the GPU, as it processes only one token per step and incurs frequent memory bandwidth bottlenecks from KV cache accesses. 
The unbalanced computation characteristic between prefill and decode stage may impose inefficiency in LLM inference \cite{DBLP:conf/osdi/AgrawalKPMKGTR24,DBLP:conf/isca/PatelCZSGMB24,DBLP:conf/osdi/ZhongLCHZL0024}.

\textbf{Scheduling Policies.}
Traditional inference engine like FasterTransformer \cite{FasterTransformer} employs batch-level scheduling which selects a group of requests and executes it until the completion of all the sequences. 
Without considering variable length characteristic of transformer architecture, this method delays early-finished and late-joining requests. 
Orca \cite{DBLP:conf/osdi/YuJKKC22} overcomes this issue by proposing iteration-level scheduling, which allows requests to dynamically enter or exit a batch before each model forward. 
\gty{Whereas}{Although} Orca can batch requests from both prefill and decode stages, it \gty{introduces}{can introduce} generation stalls for ongoing decode requests due to high prefill computation latency. 
To solve the problem, \CHUNKED~\cite{DBLP:conf/osdi/AgrawalKPMKGTR24} allows computing large prefills in small chunks across several iterations and hybrid scheduling of chunked prefill and decode tokens \gty{}{within a fixed token budget}to achieve stall-free batching. 
Specifically, \CHUNKED~first schedules all decode \gty{tokens}{requests}, then maximizes chunked prefill tokens \gty{}{scheduling}within the fixed token budget.
However, in actual situation, decode tokens often lack the opportunity to mix with adequate prefill tokens.

\subsection{Parallelism Strategies of LLM Inference}

\begin{figure}[h]
    \centering
    \includegraphics[width=\linewidth]{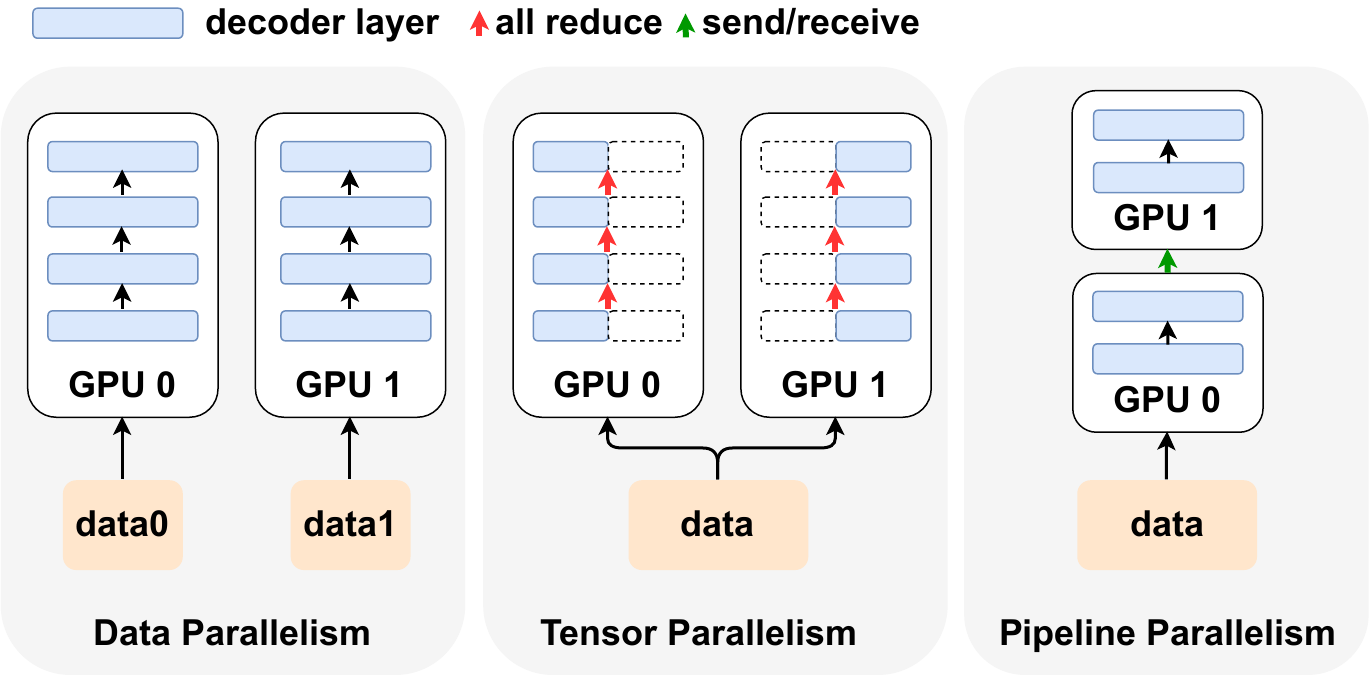}
    \caption{Comparison of data parallelism, tensor parallelism and pipeline parallelism.}
    \label{fig:dp_pp_tp}
\end{figure}

The basic parallelism strategies in LLM inference primarily consist of data parallelism and model parallelism (as shown in Figure \ref{fig:dp_pp_tp}). 
Data parallelism splits the input data into multiple parts, sending each part to the corresponding GPU for parallel processing. 
Model parallelism partitions the model itself, with each GPU responsible for a different portion of the model. 
Model parallelism can be further classified into tensor parallelism and pipeline parallelism, depending on how the model is partitioned. 
Tensor parallelism employs intra-layer parallelism, splitting individual operations across GPUs and requiring frequent communication to synchronize results. 
Pipeline parallelism \gty{adopts}{uses} inter-layer parallelism, assigning different layers to different GPUs and only requiring communication to pass intermediate activations between stages. 
In addition, pipeline parallelism employs multiple micro-batches to saturate GPUs at different pipeline stages.
Pipeline depth refers to the number of sequential stages in a pipeline, where each stage performs a specific part of a task.
Due to these differences, tensor parallelism can reduce forward latency at the cost of higher communication overhead than pipeline parallelism. 
In online serving scenarios, tensor parallelism is more suitable for low request rates, while pipeline parallelism better handles high-throughput demands.

% \subsection{Dilemma in Distributed Serving LLMs}

% % Qwen32B 4xA100
% %     cross-node  intra-node
% % TP     82.5%      66.2%
% % PP     13.7%      12.8%

% As LLMs continue to scale, single-node GPU systems often fail to accommodate entire models due to memory constraints, necessitating the adoption of model parallelism strategies. 
% Current approaches primarily include tensor parallelism and pipeline parallelism, each with distinct operational characteristics. 
% Tensor parallelism employs intra-layer partitioning, requiring continuous synchronization between devices to execute individual operations. 
% When serving 32B model with 4 GPUs (without NVLink), the communication overhead accounts for 66.2\% and 82.5\% (12.8\% and 13.7\% in pipeline parallelism) of total computation time for intra-node and cross-node deployments respectively. 
% This substantial communication cost renders tensor parallelism particularly inefficient for LLM deployments. 
% Consequently, pipeline parallelism emerges as the preferred deployment strategy due to its lower communication requirements. 
% However, this method introduces its own challenge: pipeline bubbles caused by unbalanced inter-batch computations. 
% These idle periods are recognized as efficiency limitation.

\subsection{Challenges in Pipeline Parallelism}

\begin{figure}[h]
    \centering
    \includegraphics[width=\linewidth]{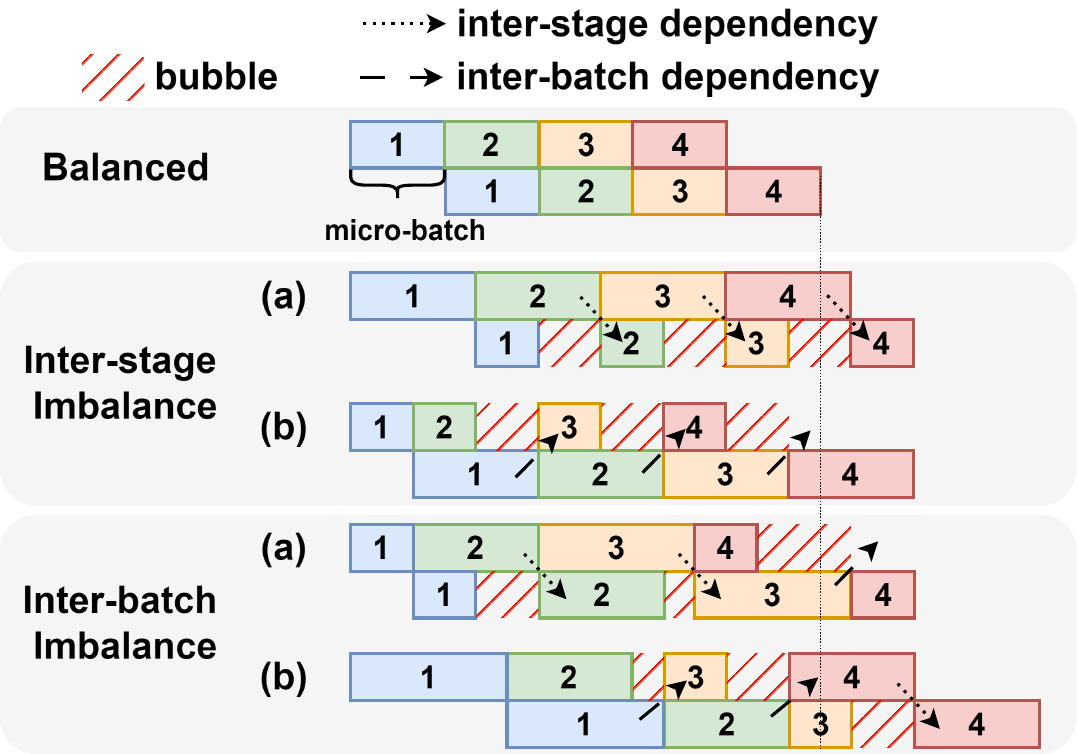}
    \caption{Pipeline bubbles caused by two types of imbalance, i.e., inter-stage and inter-batch. (the number indicates the ordinal position of the micro-batch)}
    %\caption{Pipeline bubbles caused by two types of imbalance. There are two situations (a) and (b) for each type of imbalance. The number indicates the ordinal position of the micro-batch.}
    \label{fig:pipeline_bubble}
\end{figure}

\textbf{Pipeline Bubbles.} Pipeline Parallelism reduces the need for high-bandwidth communication but may incur load-balancing issues and low GPU utilization. 
These issues manifest as pipeline bubbles, periods of GPU idle time caused by two types of dependencies: (1) inter-stage dependency, where a stage cannot begin computation until the preceding stage completes, and (2) inter-batch dependency, where the number of concurrent micro-batches is limited by the pipeline depth. 
The load imbalance stems from: (1) inter-stage imbalance due to uneven computation distribution across pipeline stages, and (2) inter-batch imbalance caused by variation in computation requirements \gty{across}{between} different micro-batches. 
Figure \ref{fig:pipeline_bubble} illustrates how these imbalances create pipeline bubbles. 
In this paper, we focus on solving inter-batch pipeline bubbles, while the inter-stage bubbles are left for future works.

\begin{figure}[ht]
    \centering
    \begin{subfigure}[b]{\linewidth}
        \centering
        \includegraphics[width=\linewidth]{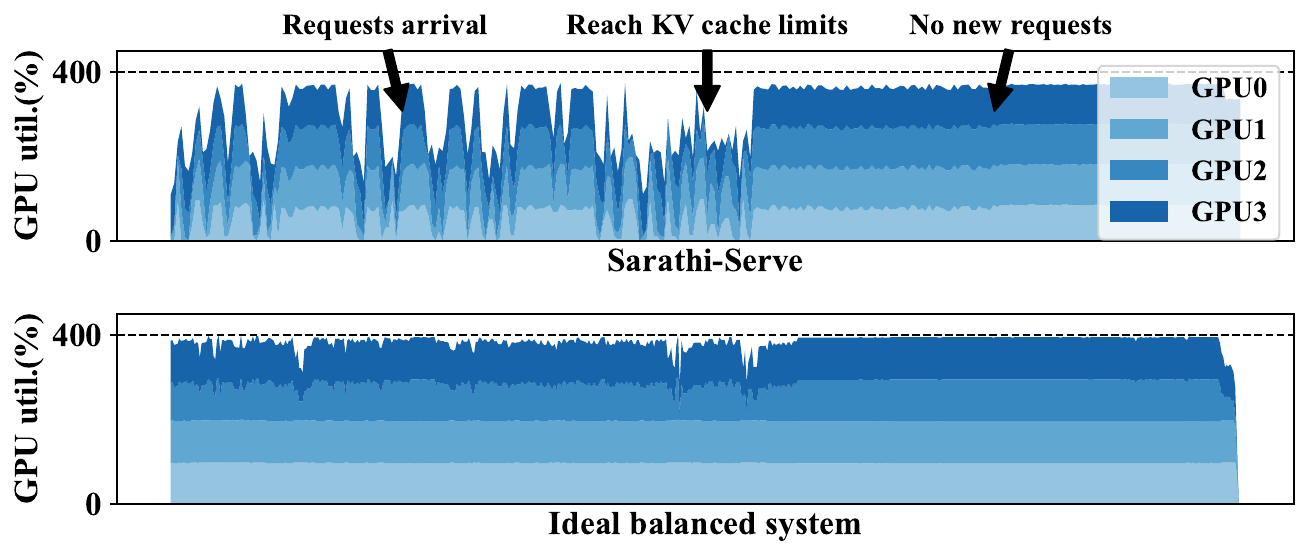}
        \caption{Variation of GPU utilization}
    \end{subfigure}
    \begin{subfigure}[b]{\linewidth}
        \centering
        \includegraphics[width=\linewidth]{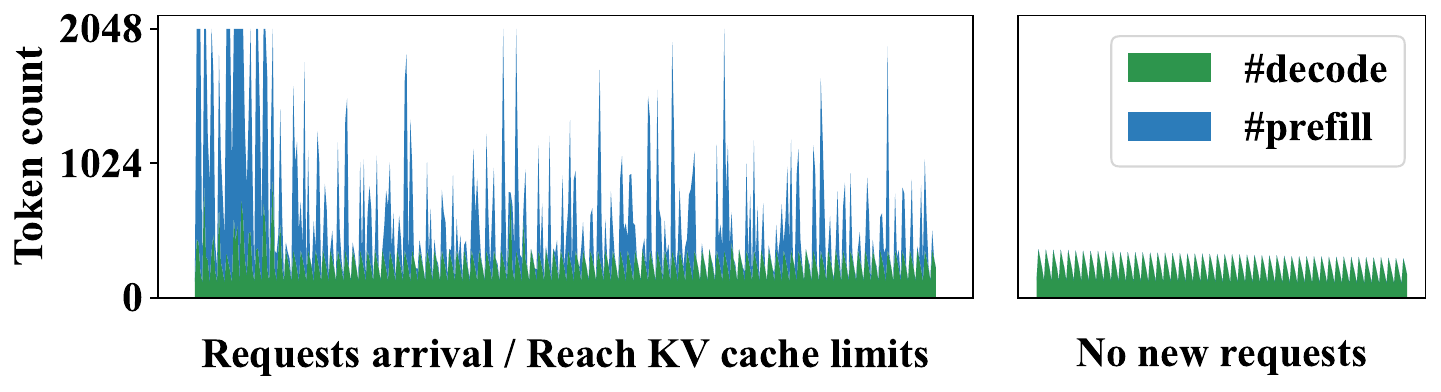}
        \caption{Batched token count in \CHUNKED}
        \label{fig:sub_token_count}
    \end{subfigure}
    \caption{Under-utilized GPU caused by unbalanced scheduling.}
    \label{fig:gpu_util}
\end{figure}

\textbf{Insufficient GPU Utilization.} Pipeline parallelism has long been suffered from low GPU utilization.
Although \CHUNKED~attempts to mitigate this issue, under-utilization persists in current systems.
As shown in Figure \ref{fig:gpu_util} when serving a 32B model with 4 GPUs, their utilization follows a two-stage pattern: an initial phase with high fluctuations followed by a stable but suboptimal phase. 
By correlating these observations with request timing, we see that the initial stage coincides with incoming requests, forcing the system to handle both prefill and decode tokens. 
Once no new requests arrive, the system shifts to decoding only, leading to steadier but still inefficient utilization.
Notably, batched token counts (Figure \ref{fig:sub_token_count}) fluctuate throughout execution, particularly upon requests arrival or reaching KV cache limits. 
These irregular batch sizes introduce severe pipeline bubbles, which directly contribute to poor GPU utilization in both stages.

\subsection{Scheduling Demands}

\begin{figure}[ht]
    \centering
    \includegraphics[width=\linewidth]{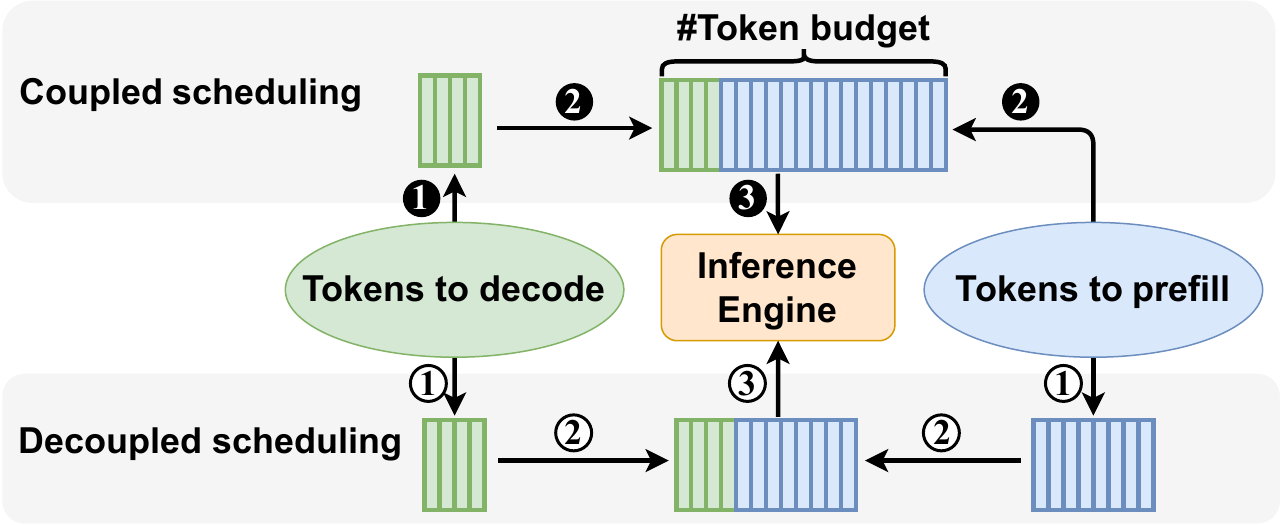}
    \caption{Comparison between coupled and decoupled scheduling.}
    \label{fig:decoupled_schedule}
\end{figure}

\textbf{Balanced Scheduling.} 
Pipeline parallelism relies on the balanced computation across micro-batches, but current scheduling strategy fails to meet this requirement:
(1) Prefill Imbalance.
The fluctuations of batched prefill tokens count within the current scheduling strategy primarily stems from two key factors.
First, prefill tokens depend on waiting requests. When no requests are available to prefill, the batched token count fluctuates. 
Second, prefill operations are constrained by KV cache utilization. 
If insufficient space exists to store the computed KV cache, the system halts the scheduling of prefill tokens. 
However, current scheduling method fails to account for these factors, resulting in unbalanced prefill \gty{scheduling}{tokens count}. 
(2) Decode Imbalance.
To achieve balanced processing during the decode stage, we aim to distribute the total decode requests as evenly as possible across all available micro-batches. 
While current scheduling strategy lacks explicit balancing considerations and may induce uneven workload distribution across micro-batches.

\textbf{Decoupled Scheduling.}
As shown in Figure \ref{fig:decoupled_schedule}, current method first schedules all decode tokens \ding{202}, then maximizes chunked prefill tokens within the fixed token budget \ding{203}.
However, we should separately schedule balanced count of prefill and decode tokens \ding{192}, rather than limiting their total to a fixed value.
This is because the scheduled tokens count may not reach the maximum budget and often causes the unbalanced schedule.
Besides, the optimal batch tokens count tends to vary dynamically.
Moreover, the tight coupling between prefill and decode scheduling often leads to interference between these stages. 
For instance, when numerous decode requests are in progress, the available token capacity for prefill becomes limited.
Nevertheless, maximizing inference throughput requires processing large batches of prefill tokens. 
This fundamental conflict reveals that tightly coupling prefill and decode scheduling under a fixed total token budget cannot effectively satisfy their respective requirements. 
Thus there is a critical need to develop decoupled scheduling mechanisms.

\textbf{Dynamically Scheduling.}
The number of prefill tokens per batch should adapt dynamically to the inference system's state. 
For example, at low KV cache utilization (i.e., when few requests are in decode), we should increase prefill speed to maximize GPU utilization. 
Conversely, during periods of high KV cache utilization, we are expected to reduce the prefill rate to prevent preemption of sequences for insufficient KV cache. 
Another concern is that if there are few tokens to prefill, we \gty{should}{want to} prefill \gty{smoothly}{slowly} to avoid sudden fluctuation in batched tokens count. 
In the contrast, if there are \gty{abundant}{many} tokens to prefill, we \gty{are expected}{want} to maintain high prefill rate.
However, current prefill batching is constrained by the fixed token budget and the number of active decode requests, rather than adapting to actual demand. 
This inflexibility leads to suboptimal scheduling policy, highlighting the need for a more adaptive scheduling approach.

%% file: chapters/4.design.tex
\section{DESIGN}

\begin{figure}
    \centering
    \includegraphics[width=\linewidth]{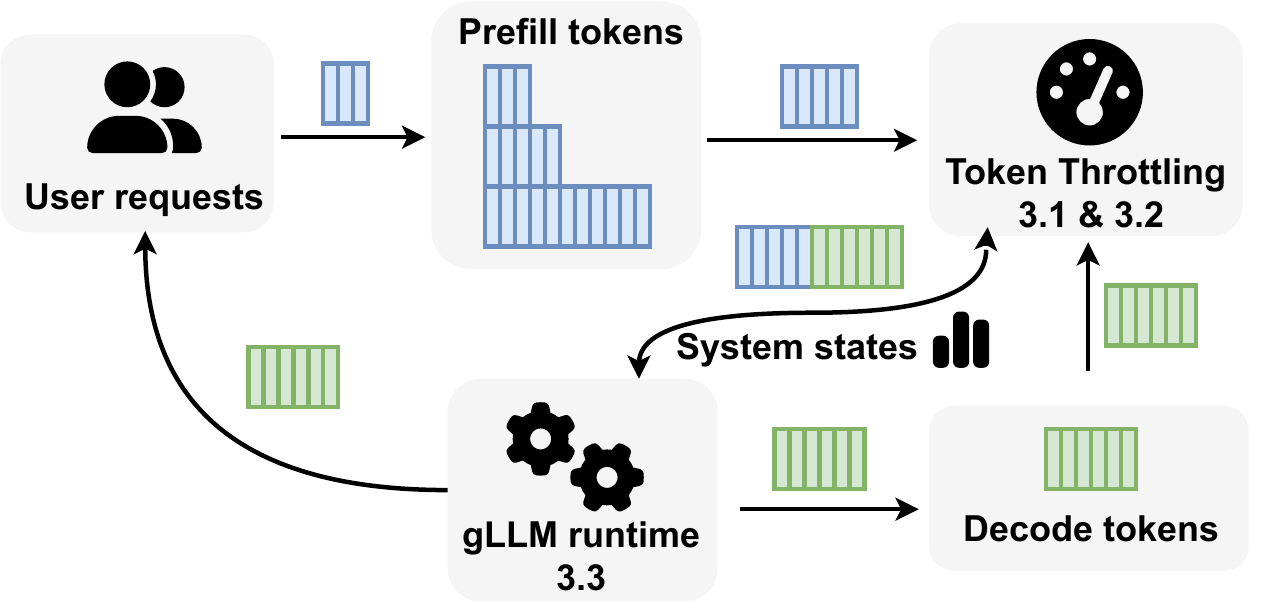}
    \caption{Overview of \NAME~system. Inference procedure in \NAME~consists of tokens flowing through its components.}
    \label{fig:overview}
    \vspace{-0.1in}
\end{figure}

\begin{figure*}[ht]
    \centering
    \includegraphics[width=\linewidth]{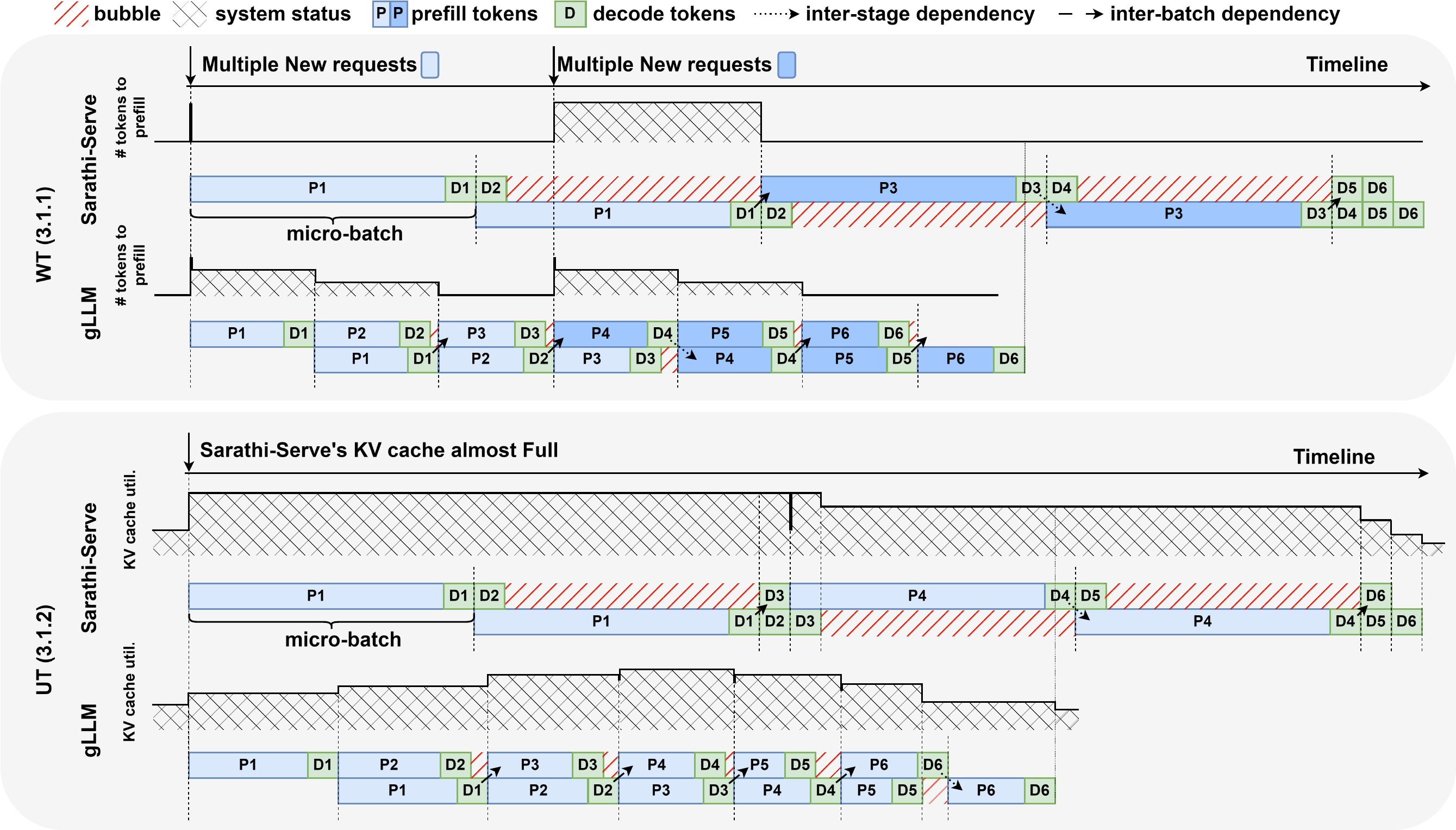}
    \caption{Case study (the pipeline depth is 2) of prefill \METHODT. The number indicates the ordinal position of the micro-batch. For the second situation, the KV cache is allocated for prefill tokens prior to the execution of each micro-batch (at the begin of first stage). After processing a micro-batch (at the end of last stage), the KV cache usage may increase, decrease or stay roughly the same depending on how many requests meet the termination criterion. The KV cache usage is consistent across all GPUs since they share unified page tables.}
    \label{fig:prefill_throttling}
\end{figure*}

In this section, we present \NAME, a global balanced pipeline parallelism systems with \METHODT. 
As shown in Figure \ref{fig:overview},
\NAME~dynamically schedules prefill and decode tokens separately by throttling batched token counts according to real-time inference system states. 
After scheduling, \NAME~merges scheduled prefill and decode tokens into a single batch. 
In addition, \NAME~\gty{runtime}{} adopts asynchronous architectures, hiding CPU operations overhead.

\subsection{Prefill \METHODT}

The prefill operation is the first step in LLM inference, where the prompt's KV cache is computed and the first output token is generated. 
Therefore, prefill scheduling depends primarily on two factors: the number of tokens waiting for prefill and the system's KV cache utilization. 
First, if no requests are pending prefill, the operation cannot be scheduled. 
Second, if the KV cache is near capacity, prefill cannot proceed due to insufficient memory. 
Beyond these constraints, the scale of these factors also influences scheduling decisions. 
When few requests are waiting or KV cache usage is high, the system should reduce the prefill rate to avoid pipeline bubbles or requests preemption. 
Conversely, when many requests are queued or KV cache availability is high, the system should increase the prefill rate to maximize throughput. 
To balance these \gty{factors}{dynamics}, we throttle the prefill token count based on both the volume of pending tokens and current KV cache pressure. 

\subsubsection{\underline{T}hrottling by Tokens Count A\underline{w}aiting Prefill (WT)}
\label{sec:prefill_factor_wait}

During each schedule, \NAME~collects the number of tokens a\underline{w}aiting \underline{p}refill ($\#WP$) to determine the batched prefill token count ($\#P$). 
This decision relies on three hyperparameters, minimum/maximum batched token count of prefill ($\#MinP$/$\#MaxP$) and the number of i\underline{t}erations ($\#T$) required to process all tokens waiting for prefill. 
The batched prefill token count can be calculated as follows:

\begin{equation}
    \#P=min(max(\frac{\#WP}{\#T},\#MinP),\#MaxP)
\end{equation}

\subsubsection{\underline{T}hrottling by KV Cache \underline{U}tilization Rate (UT)}
\label{sec:prefill_factor_KV}

At the beginning of each schedule, \NAME~also collects the KV cache idle rate ($KV_{free} \in [0,1]$) to determine batched token count ($\#P$). This decision depends on two hyperparameters, minimum/maximum tokens count batched in prefill ($\#MinP$/$\#MaxP$). The batched token count is then computed as:

\begin{equation}
    \#P=max(\#MaxP\times KV_{free},\#MinP)
\end{equation}

\subsubsection{Threshold}
\label{sec:prefill_threshold}
Besides dynamically adjusting prefill tokens count, we also introduce a KV cache idle threshold ($KV_{thresh} \in [0,1]$) to regulate scheduling decisions.
When current KV cache idle rate is less than $KV_{thresh}$, the system automatically suspends prefill token processing to prevent KV cache overflow. This safeguard mechanism addresses the critical issues observed in unrestricted prefill operations: Premature preemption of ongoing decode requests causes costly recomputation time. By maintaining buffer headroom through the threshold mechanism, we ensure adequate resource allocation for active decode requests while maintaining stable system.

Combining aforementioned factors (WT, UT and threshold), we compute the batched token count as:
\begin{equation}
    \#P=max(min(\frac{\#WP}{\#T},\#MaxP\times \frac{KV_{free}-KV_{thresh}}{1-KV_{thresh}}),\#MinP)
\end{equation}

\subsubsection{Case Study}

Figure \ref{fig:prefill_throttling} compares the prefill scheduling strategies of \CHUNKED~and \NAME. 
In the first scenario, there are no tokens waiting to prefill at the initial state and $\#T$ is set to 3.
\gty{Upon requests arrival,}{} \CHUNKED~eagerly processes prefill tokens. 
This leads to no prefill tokens batched in the following batch, resulting in large fluctuation in the computation.
The significant pipeline bubbles also prevent \CHUNKED~from processing the next batch of requests in a timely manner.
Instead, \NAME~evenly distributes new requests across the following batches, achieving better load balance.

For the second situation, there are sufficient tokens waiting to prefill at the start.
\CHUNKED~schedules prefill tokens without considering KV cache utilization rates.
This results in the remaining KV cache space being insufficient to support prefill computation after allocating the \gty{room}{KV cache} to the prefill tokens.
And the following batch can only handle decode tokens until some requests finish. 
In the contrast, \NAME~dynamically adjust prefill scheduling based on KV cache occupancy, ensuring more balanced computation.
\gty{Moreover}{}, \NAME~can achieve more efficient KV cache turnover, enabling timely processing of decode requests and preventing their KV cache from occupying space for extended periods.

\begin{figure*}[ht]
    \centering
    \includegraphics[width=.9\linewidth]{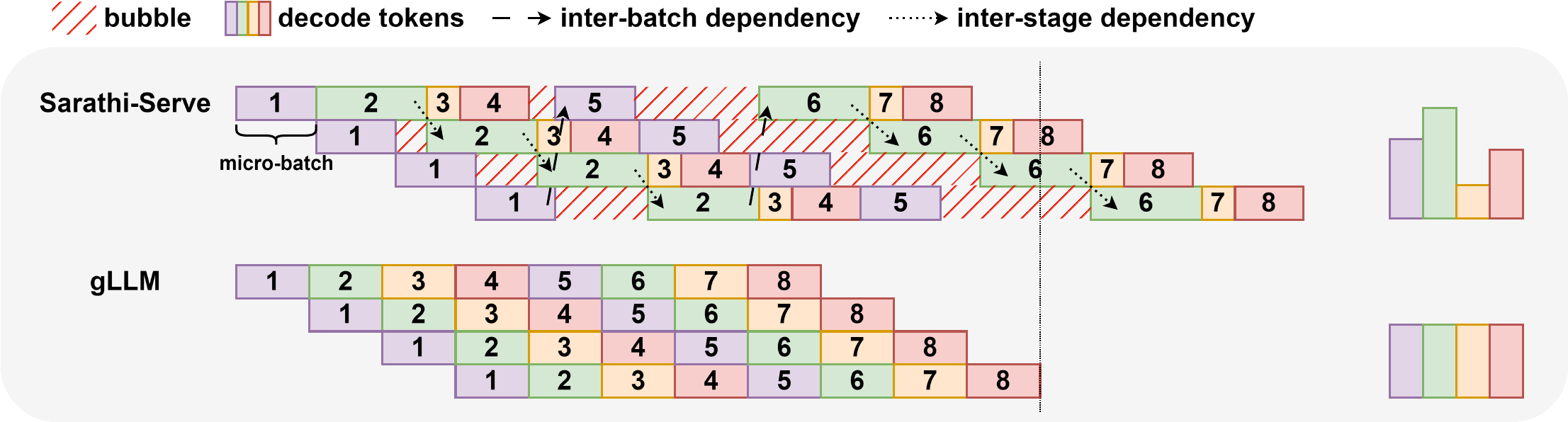}
    \caption{Case study (the pipeline depth is 4) of decode \METHODT. The number indicates the ordinal position of the micro-batch.}
    \label{fig:decode_throttling}
\end{figure*}

\subsection{Decode \METHODT}

\subsubsection{Throttling by Tokens Count Under Decode} The scheduling for decode \METHODT~is straightforward, since decode operations require multiple iterations (equal to the output sequence length) while prefill operations typically complete in a single iteration. The variation in decode requests is relatively small, as these requests either originate from completed prefill phase or have reached their termination condition. Therefore, our objective is to distribute the total decode tokens evenly across all available micro-batches. Given that the maximal number of active micro-batches equals the pipeline depth ($\#PP_{depth}$), we can compute the batched decode tokens count ($\#D$) as:
\begin{equation}
    \#D=\frac{\#RD}{\#PP_{depth}}
\end{equation}
where $\#RD$ is total \underline{r}unning tokens count under \underline{d}ecode. If the remaining decode tokens are fewer than $\#D$, we schedule all of them; otherwise, we schedule exactly $\#D$ tokens.

\subsubsection{Case Study} Figure \ref{fig:decode_throttling} compares the decode scheduling strategies of \CHUNKED~and \NAME. 
\CHUNKED~fails to balance workloads evenly across micro-batches, leading to significant pipeline bubbles.
For inter-stage dependency, the second/sixth micro-batch has to wait for the completion of the previous stage.
For inter-batch dependency, the fifth/sixth micro-batch has to wait for the finish of the first/second micro-batch.
These dependencies \gty{induce}{create} significant pipeline bubbles.
In contrast, \NAME~optimizes scheduling by dynamically adjusting token counts based on the total tokens under decode, ensuring a balanced workload distribution and reducing pipeline inefficiencies.

\subsection{\NAME~Runtime}

\begin{figure}[h]
    \centering
    \includegraphics[width=\linewidth]{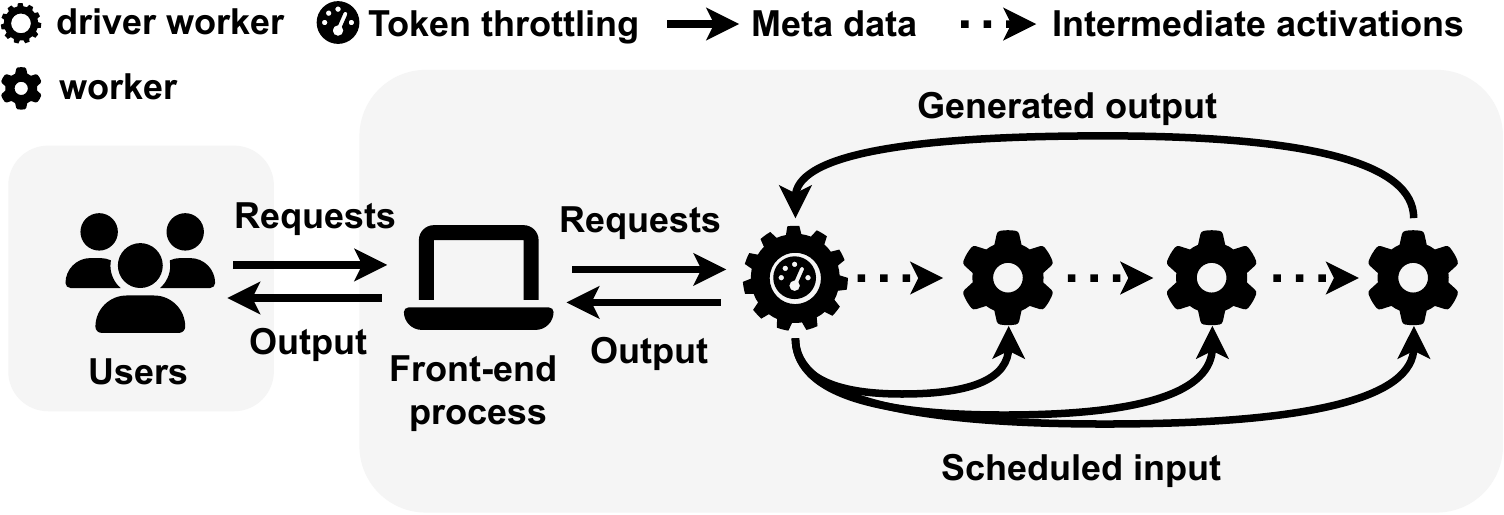}
    \caption{Architecture of \NAME~runtime.}
    \label{fig:gllm_arch}
\end{figure}

Compared to tensor parallelism, pipeline parallelism involves more sophisticated control flow to orchestrate micro-batches across different pipeline stages. 
To enable \METHODT, we design an asynchronous runtime with its architecture shown in Figure \ref{fig:gllm_arch}.

To support concurrent execution of multiple stages, \NAME~runtime adopts a multi-process architecture where each pipeline stage is assigned a dedicated worker process, along with a separate front-end process for user interaction.
The workers are divided into two roles: a driver worker and ordinary workers. 
The driver worker oversees other ordinary workers, handling tasks such as receiving new requests from the front-end, scheduling micro-batches, broadcasting metadata for each schedule and streaming output back to the front-end.
Meanwhile, all the workers focus on model execution, receiving activations from the previous stage, performing forward computations, and sending activations to the next stage. 
The driver worker is responsible for the KV cache management and all the workers share the page tables like \VLLM.

The asynchronous of \NAME~runtime is achieved through three coordinated design principles:

(1) Non-blocking pipeline operations. All workers processes employ non-blocking \gty{mechanisms}{implementations} for core operations including request reception, metadata exchange, and activation transmission, forming a continuous processing pipeline that eliminates idle waiting between computational stages.

(2) Decoupled frontend-backend processing. We \gty{design}{implement} a dedicated frontend process that handles user-facing operations (request intake and response streaming), enabling full parallelism with backend model execution on worker processes. This separation allows continuous user interaction while maintaining uninterrupted model computation.

(3) Preemptive metadata scheduling. The \gty{runtime utilizes}{system implements} a dual-phase data transmission\footnote{In \NAME, metadata is transmitted via ZeroMQ, while intermediate activations are exchanged using NCCL.} strategy: (a) Driver workers broadcast metadata packets to all workers (b) Workers receive corresponding intermediate data streams. 
This decoupling enables workers to perform critical path preparation (input or attention metadata tensor creation) using early-arrived metadata, effectively overlapping data preparation latency with active computation cycles. The proactive scheduling mechanism ensures computation resources remain fully utilized throughout the execution timeline.

\subsection{Implementation}

\begin{figure*}[h]
    \centering
    \begin{subfigure}[b]{.48\linewidth}
        \centering
        \includegraphics[width=\linewidth]{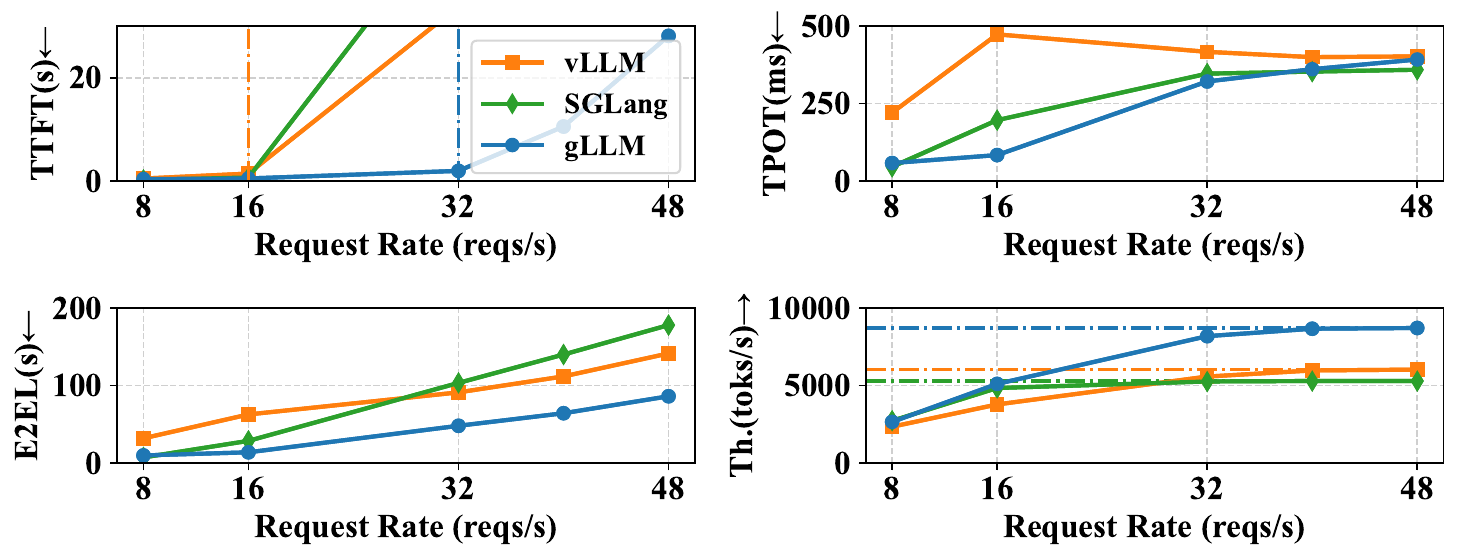}
        \caption{Qwen2.5-14B, ShareGPT}
    \end{subfigure}
    \begin{subfigure}[b]{.48\linewidth}
        \centering
        \includegraphics[width=\linewidth]{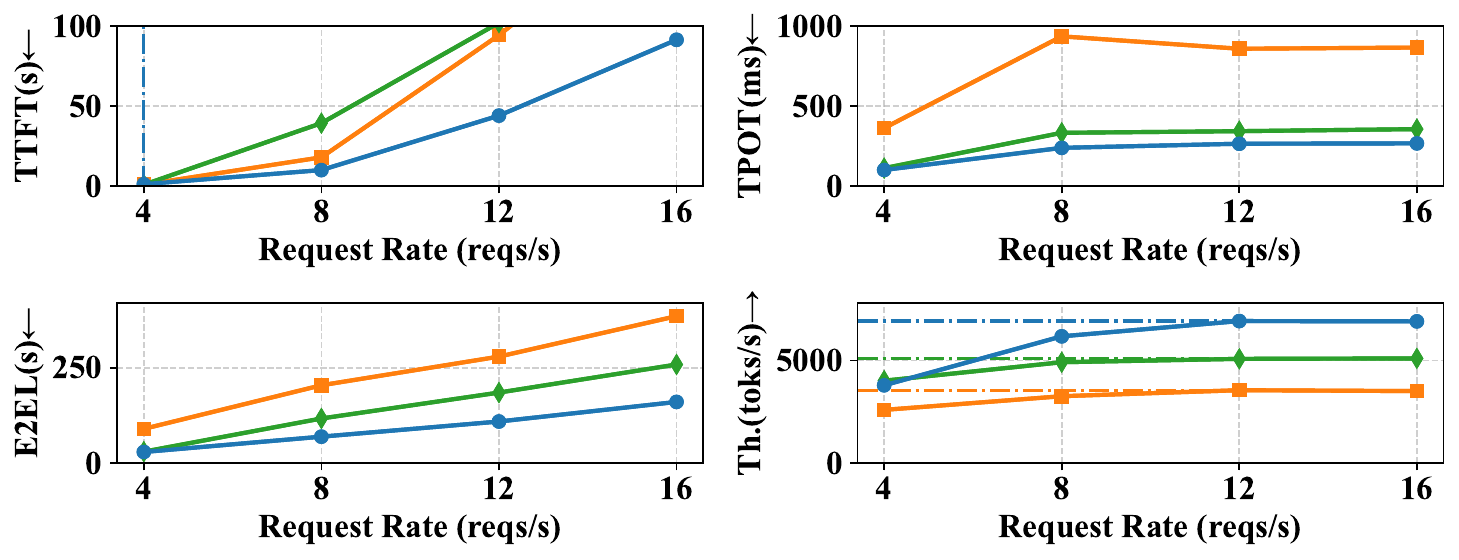}
        \caption{Qwen2.5-14B, Azure}
    \end{subfigure}
    \begin{subfigure}[b]{.48\linewidth}
        \centering
        \includegraphics[width=\linewidth]{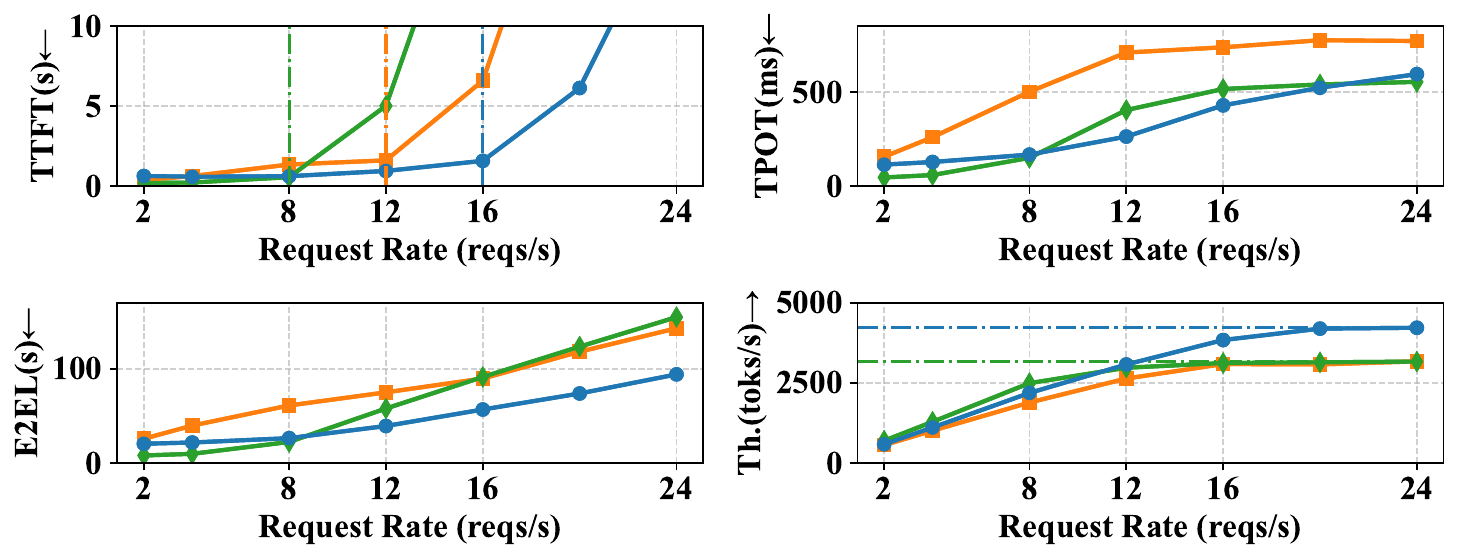}
        \caption{Qwen2.5-32B, ShareGPT}
    \end{subfigure}
    \begin{subfigure}[b]{.48\linewidth}
        \centering
        \includegraphics[width=\linewidth]{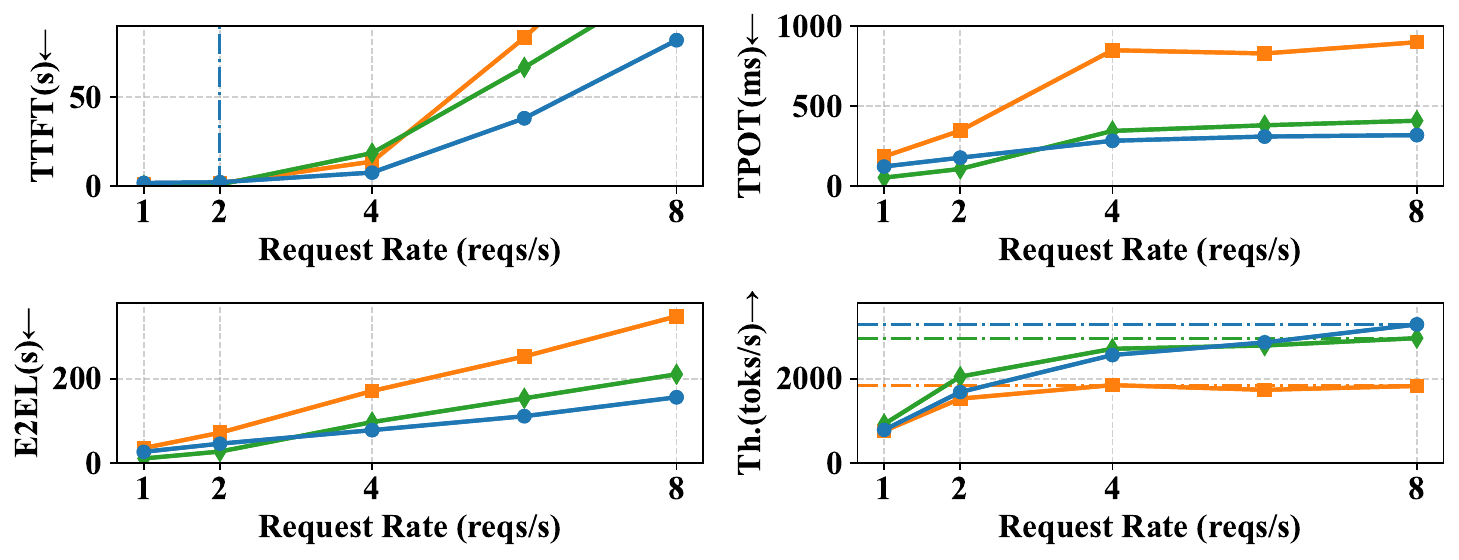}
        \caption{Qwen2.5-32B, Azure}
    \end{subfigure}
    % \begin{subfigure}[b]{\linewidth}
    %     \centering
    %     \includegraphics[width=\linewidth]{result_pic/sharegpt_Qwen72B_4_L20.pdf}
    %     \caption{Qwen2.5-72B, ShareGPT, 1$\times$4$\times$L20.}
    % \end{subfigure}
    % \begin{subfigure}[b]{\linewidth}
    %     \centering
    %     \includegraphics[width=\linewidth]{result_pic/splitwise_Qwen72B_4_L20.pdf}
    %     \caption{Qwen2.5-72B, Splitwise, 1$\times$4$\times$L20.}
    % \end{subfigure}
    \caption{The latency and throughput comparison between \VLLM, \SGLANG~and \NAME~(1 node with 4$\times$L20).}
    \label{fig:performance_intra_node}
\end{figure*}

\begin{figure}[t]
    \centering
    \begin{subfigure}[b]{\linewidth}
        \centering
        \includegraphics[width=\linewidth]{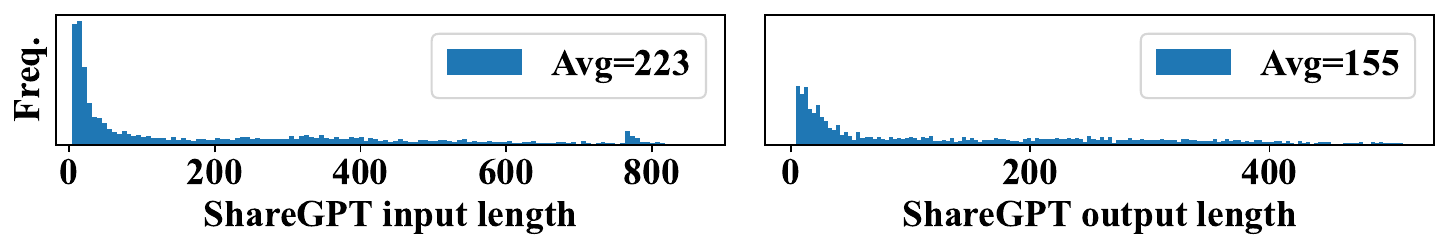}
        % \caption{ShareGPT}
    \end{subfigure}
    \begin{subfigure}[b]{\linewidth}
        \centering
        \includegraphics[width=\linewidth]{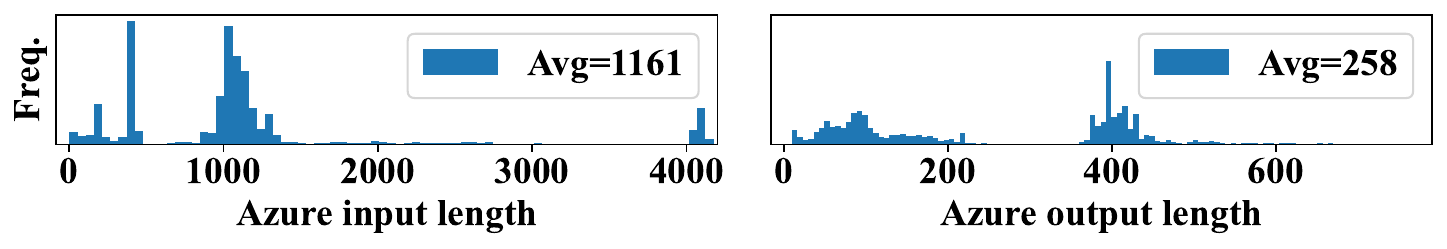}
        % \caption{Azure}
    \end{subfigure}
    \caption{Distribution of input and output length of the sampled dataset.}
    \label{fig:dataset_dis}
\end{figure}

Upon implementation, we identify a critical design limitation in \VLLM's pipeline parallelism architecture: it tightly couples the transmission of intermediate activations with input scheduling metadata. 
This integration introduces significant CPU overhead during input preparation for model forwading, accounting for approximately 17\% of the total execution time and substantially diminishing the potential benefits of pipeline parallelism. 

To overcome this limitation, we \gty{have implemented essential part of \NAME~with 4K lines of Python code on top of CUDA ecosystem}{develop \NAME, an end-to-end distributed pipeline parallelism systems designed for LLMs}. 
The system features a RESTful API frontend and offers core OpenAI-compatible APIs.
It~incorporates \VLLM's manually optimized CUDA kernels for key operations including activation functions, layer normalization, position encoding, and KV cache management, while also adopting flash attention \cite{DBLP:conf/nips/DaoFERR22,DBLP:conf/iclr/Dao24,DBLP:conf/nips/ShahBZTRD24}.
We also integrates several recent advanced LLM optimizations, including iteration-level scheduling \cite{DBLP:conf/osdi/YuJKKC22}, PagedAttention \cite{DBLP:conf/sosp/KwonLZ0ZY0ZS23}, chunked prefill \cite{DBLP:conf/osdi/AgrawalKPMKGTR24} and prefix caching \cite{DBLP:conf/nips/ZhengYXS0YCKSGB24}.
The system supports leading open-source LLMs like Llama \cite{DBLP:journals/corr/abs-2407-21783}, Qwen \cite{DBLP:journals/corr/abs-2412-15115} and ChatGLM \cite{DBLP:journals/corr/abs-2406-12793} with varying sizes. 
To validate its output quality, we evaluate \NAME~on the MMLU-pro benchmark \cite{DBLP:conf/nips/WangMZNCGRAHJLK24}, with detailed results presented in Section \ref{sec:eval_func}.

% However, we rewrite the core code of \NAME~to address significant code redundancy in \VLLM's original implementation (V0 engine), which introduced substantial CPU overhead \cite{SchedulingOverhead} compared to \SGLANG. Although \VLLM~has recently initiated a code refactoring effort (V1 engine \cite{vLLM_V1}) to resolve these issues, our evaluation shows the new engine remains unstable and requires substantial engineering effort to resolve frequent system crashes.

% Regarding architectural choices, while \SGLANG~demonstrates superior performance in tensor parallelism scenarios, its current framework lacks native support for pipeline parallelism. Adapting \SGLANG's architecture for pipeline parallelism would necessitate extensive structural modifications. To address this gap, we developed a minimal yet functional pipeline parallelism system as \NAME. Through comprehensive validation, we have demonstrated that \NAME~maintains output quality parity with \VLLM~or \SGLANG~while achieving enhanced system efficiency.

%% file: chapters/6.result_analysis.tex
\section{EVALUATION}

\subsection{Experimental Setup}

\textbf{Models and Environment.} 
We evaluate \NAME~using the Qwen2.5 series \cite{DBLP:journals/corr/abs-2412-15115} (14B and 32B parameter variants) and 
Llama-3.1-100B\footnote{This model is downscaled from Llama3.1-405B to fit in GPU memory.} \cite{DBLP:journals/corr/abs-2407-21783} for their strong multi-task performance. 
All models utilize bfloat16 data type.
Our experiments employ three node configurations: (1) 4$\times$ NVIDIA L20-48GB GPUs
(2) 4$\times$ NVIDIA A100-40GB GPUs  
(3) 4$\times$ NVIDIA A800-80GB GPUs.
All GPUs are connected via PCIe across three configurations.
We primarily evaluated two types of scenarios: intra-node deployment and cross-node deployment.
For intra-node experiments, we use the L20 configuration. 
Cross-node evaluations leverage the A100 and A800 setup with simulated network conditions achieved by disabling both P2P communication (PCIe-based) and shared memory access (by setting environment variable $NCCL\_SHM\_DISABLE=1$ and $NCCL\_P2P\_DISABLE=1$). 
This configuration forces all inter-GPU communication through the network stack. 
Testing results show that the simulated network communication bandwidth achieves 73.28 Gbps, whereas the PCIe-based communication bandwidth attains 20.79 GB/s.

\textbf{Workloads.} We synthesize workloads based on ShareGPT \cite{ShareGPT} and Azure \cite{DBLP:conf/isca/PatelCZSGMB24}, which both comes from real LLM services. The ShareGPT dataset is a collection of user-shared conversations with ChatGPT. The Azure dataset is from production traces taken from Azure LLM inference services, including the arrival time, input size and output size. 
Figure \ref{fig:dataset_dis} displays the distribution of input and output lengths across the sampled datasets, revealing that the Azure dataset has a notably longer average input (5.21$\times$) and output (1.66$\times$) length compared to ShareGPT.
We mimic the cloud service scenario and generate request arrival times using Poisson distribution with different request rates.

\textbf{Schemes.} To evaluate the effectiveness of our proposed design, we compare \NAME~against the following systems\footnote{We do not add prefill-decode disaggregated architectures to the baselines because \NAME~can be applied into it as a superior backend system.}. 
\begin{itemize}
    \item \textbf{\VLLM} \cite{DBLP:conf/sosp/KwonLZ0ZY0ZS23}. We leverage \VLLM~(v0.8.1) as the fastest inference engines for pipeline parallelism. The framework offers two backend engine versions, V0 and V1. While V1 demonstrates superior speed to V0, our testing revealed it to be less stable. As a result, we employ V1 as the default choice but fall back to V0 when encountering engine crashes.
    \item \textbf{\SGLANG} \cite{DBLP:conf/nips/ZhengYXS0YCKSGB24}. We leverage \SGLANG~(v0.4.3.post2) as the most efficient inference engine for tensor parallelism implementations. While \SGLANG~has lower CPU overhead than \VLLM, it currently lacks pipeline parallelism support.
    \item \textbf{\NAME}. Proposed global balanced pipeline parallelism systems with \METHODT.
    \item  \textbf{\NAME~w/o WT}. \NAME~without WT (section \ref{sec:prefill_factor_wait}).
    \item \textbf{\NAME~w/o UT}. \NAME~without UT (section \ref{sec:prefill_factor_KV}).
    \item \textbf{\NAME~w/ CK}. \NAME~with the scheduling policy used in \CHUNKED.
\end{itemize}
\VLLM~and \SGLANG~both follows \CHUNKED's scheduling strategy (token budget is set to 2048).
To ensure fairness comparisons, we disable KV cache reuse across requests and cuda graph for each system. 
While \VLLM~and \NAME~employ pipeline parallelism, \SGLANG~utilizes tensor parallelism.
The GPU memory utilization of each system is set to the maximum without encountering out of memory error. 
In the main experiment, we benchmark \NAME~against \VLLM~and \SGLANG, while the ablation study compares \NAME~with its variations. 
For \NAME, we set the hyperparameters as follows: $\#T=8$, $\#MaxP=2048$, $\#MinP=32$ and $KV_{thresh}=0.05$. We examine the impact of these parameter settings in the sensitivity study.

\textbf{Metrics.} We use the following metrics to measure the performance of the systems.
\begin{itemize}
    \item \textbf{Time to first token (TTFT)}: Average time taken from when a user sends a prompt to the LLM until the first token of the response is generated.
    \item \textbf{Time per output token (TPOT)}: Average time required to generate each subsequent token after the first one.
    \item \textbf{End to end latency (E2EL)}: Average total time from prompt submission to the completion of the full response.
    \item \textbf{Throughput}: Average input and output tokens processing throughput.
    \item \textbf{SLO Attainment}: The SLO fulfillment rate under the given TTFT and TPOT constraints.
\end{itemize}

\begin{figure}[ht]
    \centering
    \vspace{-0.1in}
    \begin{subfigure}[b]{\linewidth}
        \includegraphics[width=\linewidth]{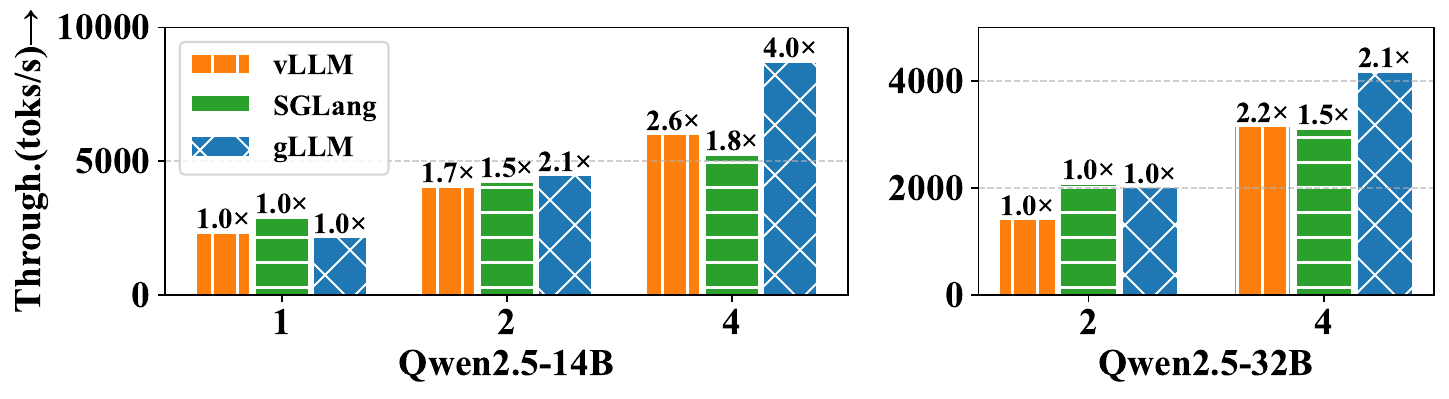}
        \caption{Intra-node scalability with L20}
    \end{subfigure}
    \begin{subfigure}[b]{\linewidth}
        \includegraphics[width=\linewidth]{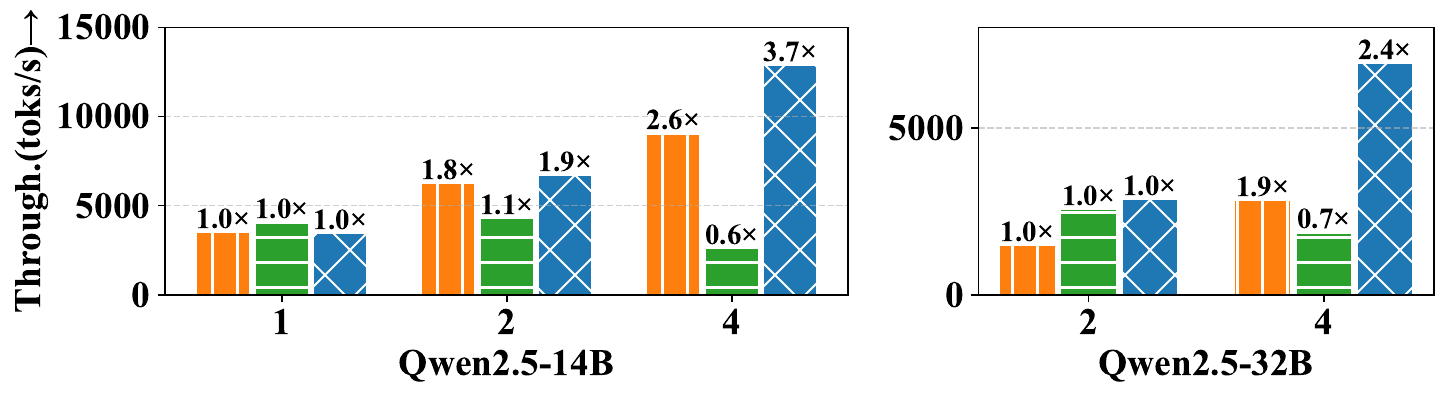}
        \caption{Cross-node scalability with 1$\times$A100 per node}
    \end{subfigure}
    \caption{Variation of maximum throughput as the number of GPUs/nodes (horizontal axis) increases. The numbers ($\times$) on the bar illustrate the multiples compared to single/two-card/node(s) performance.} 
    \label{fig:scalability}
\end{figure}

\begin{figure*}[h]
    \centering
    % \begin{subfigure}[b]{\linewidth}
    %     \centering
    %     \includegraphics[width=\linewidth]{result_pic/sharegpt_Qwen14B_4_A100.pdf}
    %     \caption{Qwen2.5-14B, ShareGPT, 4 nodes with 1$\times$A100 (40GB) per node.}
    % \end{subfigure}
    % \begin{subfigure}[b]{\linewidth}
    %     \centering
    %     \includegraphics[width=\linewidth]{result_pic/splitwise_Qwen14B_4_A100.pdf}
    %     \caption{Qwen2.5-14B, Splitwise, 4 nodes with 1$\times$A100 (40GB) per node.}
    % \end{subfigure}
    \begin{subfigure}[b]{.48\linewidth}
        \centering
        \includegraphics[width=\linewidth]{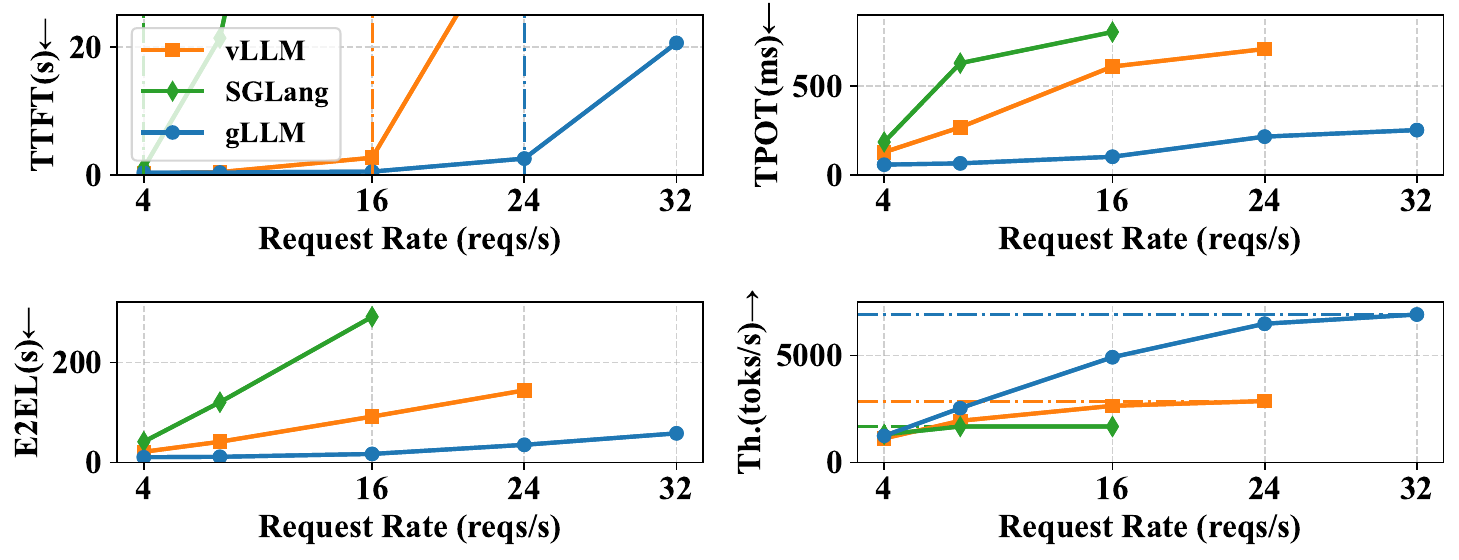}
        \caption{Qwen2.5-32B, ShareGPT, 1$\times$A100 per node}
    \end{subfigure}
    \begin{subfigure}[b]{.48\linewidth}
        \centering
        \includegraphics[width=\linewidth]{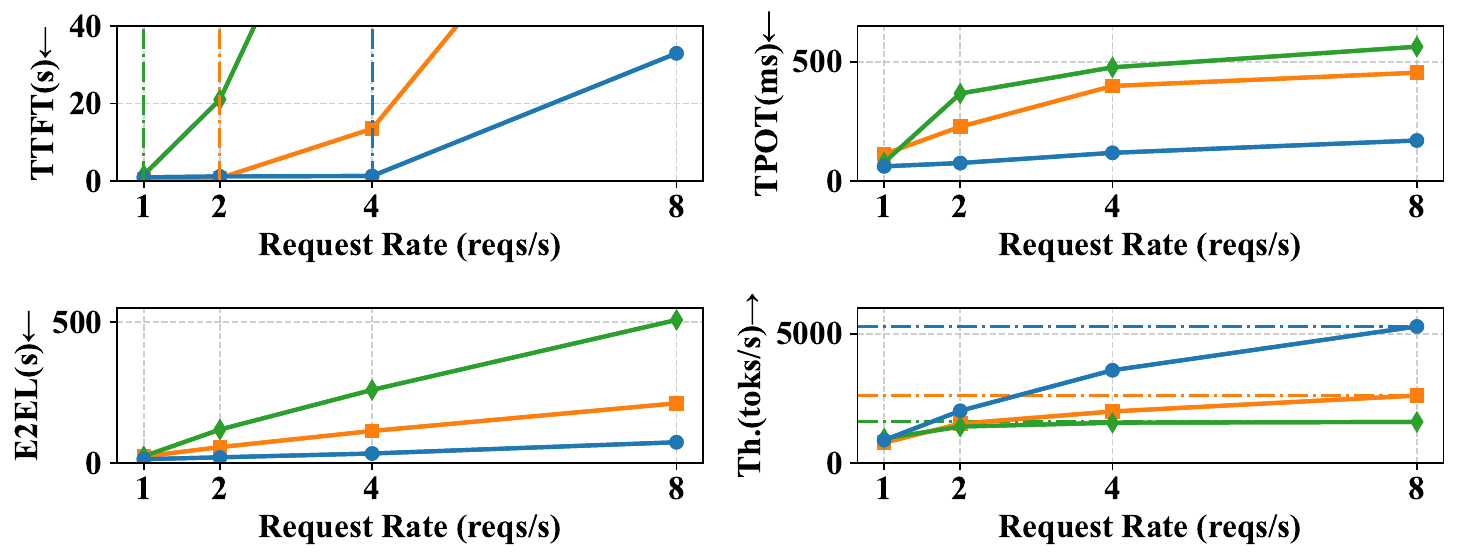}
        \caption{Qwen2.5-32B, Azure, 1$\times$A100 per node}
    \end{subfigure}
    \begin{subfigure}[b]{.48\linewidth}
        \centering
        \includegraphics[width=\linewidth]{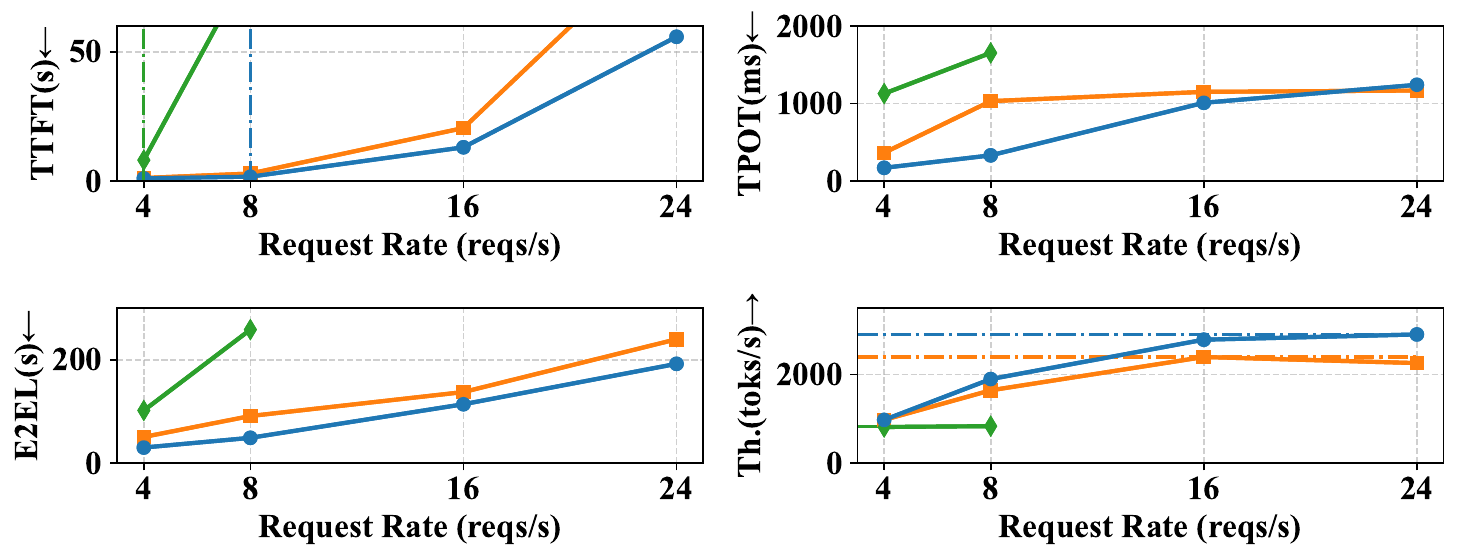}
        \caption{Llama3.1-100B, ShareGPT, 1$\times$A800 per node}
    \end{subfigure}
    \begin{subfigure}[b]{.48\linewidth}
        \centering
        \includegraphics[width=\linewidth]{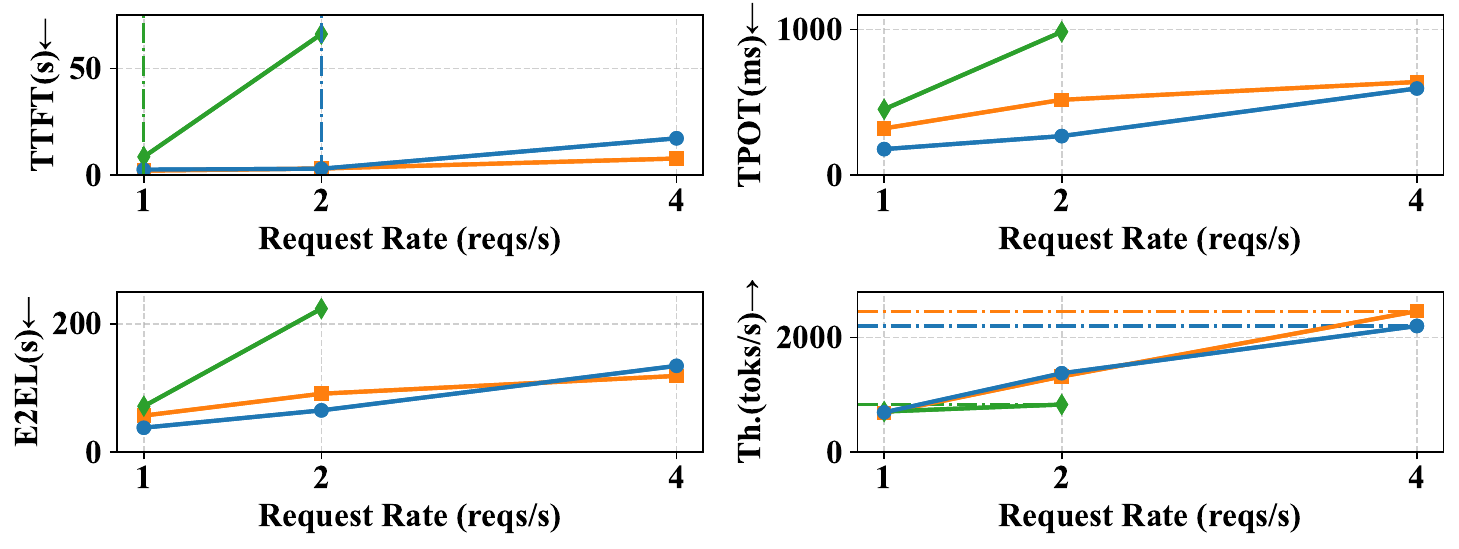}
        \caption{Llama3.1-100B, Azure, 1$\times$A800 per node}
    \end{subfigure}
    \caption{The latency and throughput comparison between \VLLM, \SGLANG~and \NAME~(4 nodes).}
    \label{fig:performance_cross_node}
\end{figure*}

\subsection{Latency and Throughput Evaluation}
To compare the performance of \VLLM, \SGLANG~and \NAME, we conduct experiments based on ShareGPT and Azure dataset using models from 14B to 100B with the results illustrated on Figure \ref{fig:performance_intra_node}\footnote{↑ indicates higher values are better, while ↓ means lower values are preferable.} (intra-node) and Figure \ref{fig:performance_cross_node} (cross-node). 
We can summarize a few key points from the figure: 
As request rates increase,
(1) latency rises while throughput gradually plateaus. 
This plateau represents the system's maximum processing capacity.
(2) TTFT will rise significantly at some point due to requests queuing. 
At the most situations, \NAME~reaches its turning point at 2-6$\times$ higher request rates compared to other systems.
(3) E2EL shows an approximately linear increase trend. 
At the most situations, \NAME~achieves 0.14-0.92$\times$ lower slope compared to the other two systems.
(4) In most cases, benefiting from the more efficient scheduling policy and runtime, \NAME~significantly outperforms \VLLM~in both latency and throughput across the tested scenarios (processing capacity improves 0.29-1.50$\times$). 
While \NAME~performs slightly worse than \VLLM~when serving Llama3.1-100B on the Azure dataset at a request rate of 4, this is due to the abundance of tokens for prefilling and sufficient memory to hold the KV cache, which minimizes variations of scheduled tokens in \CHUNKED. 
However, such conditions are uncommon in real-world serving environments.
(5) Tensor parallelism is well-suited for scenarios with low request rates and high bandwidth connection. 
For intra-node experiment, \SGLANG~achieves lower latency compared to pipeline parallelism systems under low request rates.
But as the request rate increases, the advantage of \SGLANG~diminishes and even less than \NAME.
For inter-node experiment, the performance of \NAME~is significantly better than \SGLANG~due to its high communication overhead. (processing capacity improves 0.11-3.98$\times$)
(6) Compared to Azure, \NAME~is better at handling ShareGPT. The reason is that \NAME~can balance prefill tokens across different requests. While sequences in Azure encompass longer input, weakening inter-request parallelism.

% (3) As the model size continues to increase, the advantage of \NAME~over \SGLANG~gradually diminishes or even falls behind. This is because when deploying models with larger parameters, less space is left for the KV cache, resulting in fewer requests being processed simultaneously in the pipeline and thereby reducing the pipeline's efficiency. 

% (4) \NAME~demonstrates a larger advantage on the ShareGPT dataset compared to Splitwise. This is due to Splitwise's average input length being four times longer than ShareGPT's, requiring more KV cache per request, which reduces the number of requests in the pipeline and ultimately lowers inference efficiency. 

% Based on the above analysis, pipeline parallelism demands greater GPU memory capacity than tensor parallelism to hold more concurrent inference requests. Hence, utilizing Compute Express Link (CXL) for GPU memory expansion could be a viable solution in pipeline parallel scenarios.

\subsection{Scalability Study}

To evaluate the scalability of \NAME, we conduct maximum throu-ghput tests against \VLLM~and \SGLANG~by incrementally increasing request rates until system throughput stabilizes. Our results shown in Figure \ref{fig:scalability} reveal distinct scaling patterns across systems. While \NAME~demonstrates marginally lower throughput than \VLLM~and \SGLANG~when serving 14B model on single GPU configurations (attributed to incomplete optimizations), it achieves near-linear scaling efficiency as GPU counts increase. In contrast, \VLLM~exhibits sub-linear scaling with the 14B model but maintains linear scaling with the 32B variant. \SGLANG~demonstrates sub-linear scaling within a single node, but experiences performance degradation as GPUs count increase in cross-node deployments due to high inter-GPU communication overhead. Notably, \NAME's performance advantage becomes progressively more pronounced with more GPUs, suggesting superior architectural scalability.

\subsection{SLO Attainment}

\begin{figure}[ht]
    \centering
    \begin{subfigure}[b]{\linewidth}
        \includegraphics[width=\linewidth]{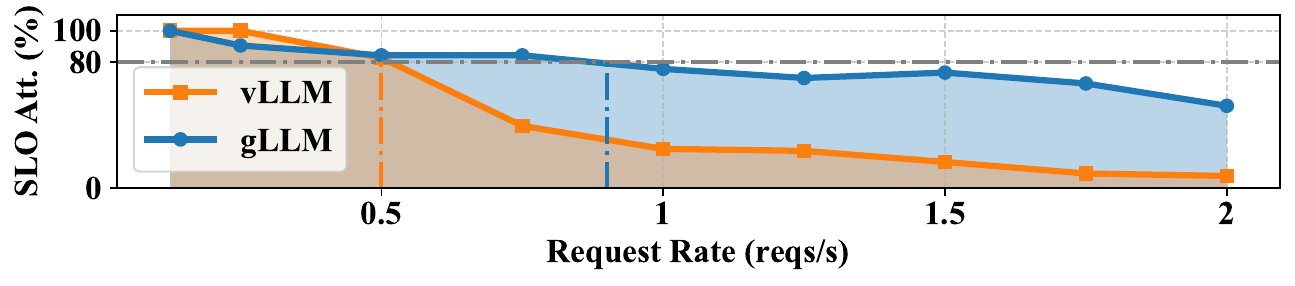}
        \caption{ShareGPT with SLO TTFT:3000ms and TPOT:150ms}
    \end{subfigure}
    \begin{subfigure}[b]{\linewidth}
        \includegraphics[width=\linewidth]{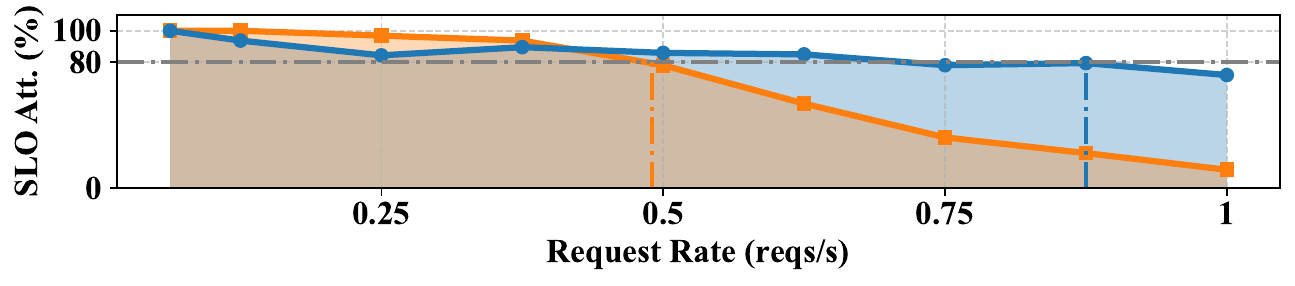}
        \caption{Azure with SLO TTFT:4000ms and TPOT:200ms}
    \end{subfigure}
    \caption{SLO attainment in cross-node deployments of Llama3.1-100B with A800.} 
    \label{fig:goodput}
\end{figure}

To evaluate SLO attainment, we compare \NAME~and \VLLM~in cross-node deployments of Llama3.1-100B. 
As shown in Figure \ref{fig:goodput}, \NAME~achieves 64\% higher SLO attainment coverage than \VLLM~across various request rates. 
Under 80\% SLO attainment, \NAME~can sustain 79\% higher request rate compared to \VLLM. 
While at low request rate, \NAME~exhibits slightly lower SLO attainment due to a marginal increase in TTFT caused by \METHODT, which occasionally exceeds the time constraint. 
To address this, we can fine-tune the hyperparameter $\#T$ to balance TTFT and TPOT performance. 

\subsection{Ablation Study}

\begin{figure}[ht]
    \centering
    \includegraphics[width=\linewidth]{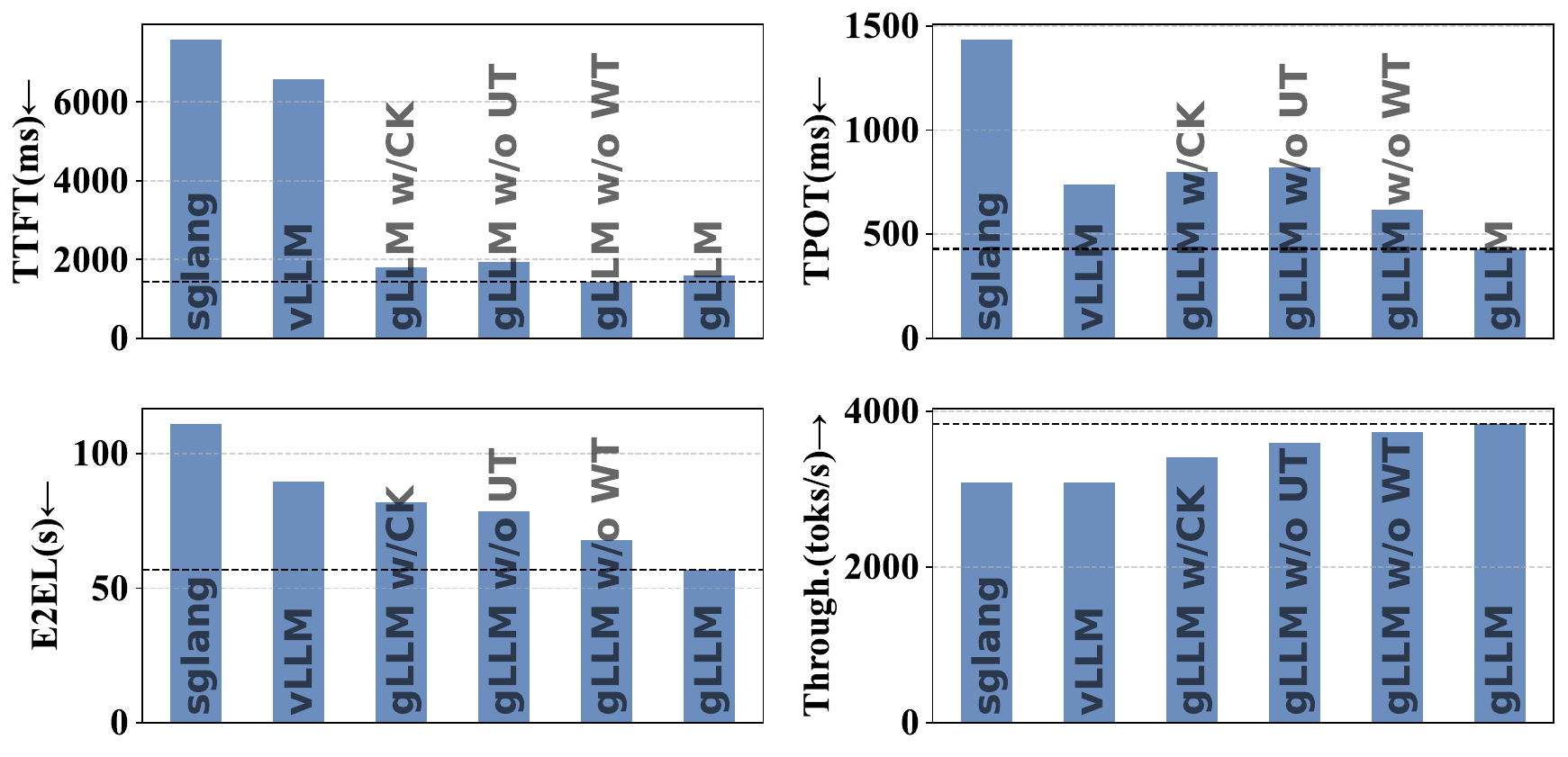}
    \caption{An ablation study on the design choices of gLLM. The dashed line marks the optimal value under each metric.}
    \label{fig:ablation}
\end{figure}

To evaluate the effectiveness of our design, we conduct ablation experiments on \NAME, with the results shown in Figure \ref{fig:ablation}. 
Notably, \NAME~w/o WT achieves 10\% lower TTFT compared to \NAME, as WT’s balanced scheduling approach slightly compromises prefill speed.
However, this comes at the cost of 44\% higher TPOT and 20\% longer E2EL.
The absence of UT leads to even more significant performance degradation, increasing TTFT by 22\%, TPOT by 91\%, and E2EL by 38\%, demonstrating UT's crucial role in balancing computation.
% Meanwhile, \NAME~w/ CK performs the worst in throughput and E2EL due to its use of basic scheduling strategy from \CHUNKED, which overlooks the balance of inter-batch computation.
Meanwhile, even with \CHUNKED's basic scheduling strategy, \NAME~w/CK achieves 10\% higher throughput and 8\% lower E2EL compared to \VLLM. 
This demonstrates that \NAME~runtime is more efficient than that of \VLLM. 

\subsection{Sensitivity Study}

\begin{figure}[h]
    \centering
    \centering
    \includegraphics[width=\linewidth]{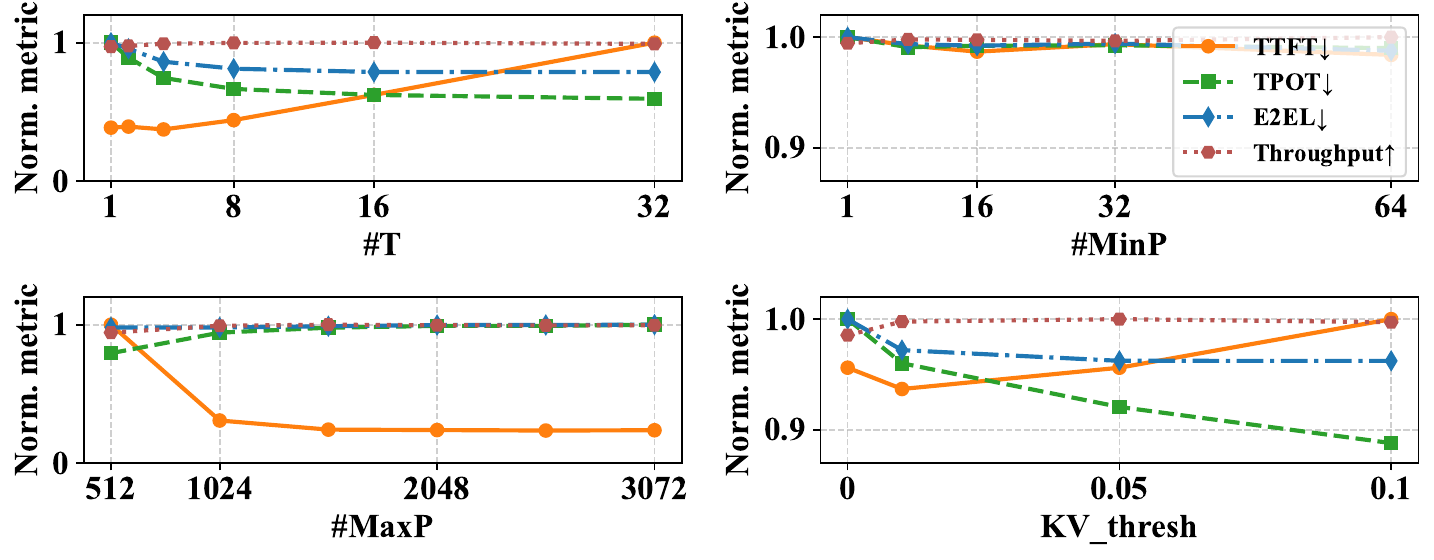}
    \caption{Normalized metric under different setting of hyperparameters \#T, \#MaxP, \#MinP and $\text{KV}_{\text{thresh}}$.}
    \label{fig:sensitivity}
\end{figure}

To examine the effect of the hyperparameters used in \NAME, we conduct a sensitivity study and the normalized results are shown in Figure \ref{fig:sensitivity}. 

\subsubsection{Impact of $\#T$}
The parameter $\#T$, representing the number of iterations to process all pending prefill tokens. 
As $\#T$ increases, the number of prefill tokens per micro-batch decreases, which generally leads to longer TTFT due to reduced prefill rate. 
However, when $\#T$ grows from 1 to 4, the TTFT remains stable because gains in computational parallelism, such as better GPU utilization and reduced idle time, offset the smaller batch sizes. 
Concurrently, TPOT improves as smaller micro-batches enable faster decoding rates. 
Throughput increases and E2EL decreases with larger $\#T$, proving \METHODT~improve overall processing efficiency through more evenly distributed computations.

\subsubsection{Impact of $\#MaxP$}
The hyperparameter $\#MaxP$ governs the maximum number of batched prefill tokens processed per scheduling iteration. 
Increasing $\#MaxP$ creates a dual effect: Higher prefill rates accelerate initial token generation (TTFT), while expanded batch sizes prolong per-token processing latency (TPOT).
However, excessively conservative $\#MaxP$ settings (e.g., 512) degrade system throughput due to suboptimal prefill rate, which constrains the system's capacity to handle concurrent requests during decode phases. 

\subsubsection{Impact of $KV_{thresh}$}

$KV_{thresh}$ defines the idle rate threshold for KV cache utilization.  
% Increasing $KV_{thresh}$ progressively reduces the prefill rate, resulting in higher TTFT and lower TPOT. 
Setting $KV_{thresh}$ to zero introduces inefficiencies: TTFT, TPOT and E2EL exhibit slight increases, while throughput declines. 
This occurs because a zero threshold pushes the system toward KV cache capacity limits more frequently, forcing preemption of ongoing requests to accommodate remaining decode requests. 
Such preemption waste computational resources and degrade overall performance.

\subsubsection{Impact of $\#MinP$}

The minimum batched prefill token count ($\#MinP$) exhibits limited impact on overall system performance across most configurations (within 2\% performance fluctuations).

\subsection{Functionality Study}
\label{sec:eval_func}

\begin{table}[ht]
    \vspace{-0.1in}
    \renewcommand\arraystretch{1.15}
    \centering
    \caption{Comparison of the number of lines of code and the score of MMLU-Pro \cite{DBLP:conf/nips/WangMZNCGRAHJLK24} (evaulated on Qwen2.5-32B-Instruct) between \NAME, \SGLANG~and \VLLM.}
    \begin{tabular}{c|c|c|c|c}
        \hline
       Framework & \NAME & \SGLANG & \VLLM(V1) & \VLLM(V0) \\ \hline \hline
       Lines of code & 3874  & 65097   & \multicolumn{2}{c}{226874} \\ \hline
       MMLU-pro↑     & 68.86 & 68.85   & NA & 69.17 \\ \hline
    \end{tabular}
    \label{tab:quality}
\end{table}

Table \ref{tab:quality} presents a comparison in lines of code and output quality across different frameworks.
Notably, \NAME~achieves comparable output quality to \VLLM\footnote{Due to the stability problem, \VLLM~with V1 crashes during the test.}~and \SGLANG~while maintaining a superior inference speed.

%% file: chapters/7.related_work.tex
\section{RELATED WORK}

\textbf{Scheduling in LLMs.} 
Traditional DNN inference frameworks primarily employ batch-level scheduling \cite{FasterTransformer}, which struggles to handle the variable sequence lengths inherent in LLMs. 
To address this limitation, Orca \cite{DBLP:conf/osdi/YuJKKC22} introduces iteration-level scheduling, enabling dynamic request admission and early exit before model execution. 
However, this approach faces challenges when processing lengthy prefill requests, as they can delay subsequent decode requests. 
To mitigate this imbalance, recent work proposes \CHUNKED~ \cite{DBLP:conf/osdi/AgrawalKPMKGTR24}, which runs prefill and decode operations together by splitting long sequences into smaller chunks. 
Despite these advancements, computational imbalance persists across each batch, significantly degrading the efficiency of pipeline parallelism.

% FasterTransformer \cite{FasterTransformer}
% Orca \cite{DBLP:conf/osdi/YuJKKC22}
% Sarathi-Serve \cite{DBLP:conf/osdi/AgrawalKPMKGTR24}

\textbf{LLMs serving systems.}
To efficiently serve LLMs, researchers have developed various systems. 
Orca \cite{DBLP:conf/osdi/YuJKKC22}, for instance, introduces a distributed serving system with iteration-level scheduling to improve throughput. 
For memory optimization, \VLLM~\cite{DBLP:conf/sosp/KwonLZ0ZY0ZS23} employs paged attention, imitating virtual memory mechanisms to reduce fragmentation, while \SGLANG~\cite{DBLP:conf/nips/ZhengYXS0YCKSGB24} leverages radix attention to eliminate redundant KV cache computations across requests. 
To address the divergent computational demands of the prefill and decode stages, Splitwise \cite{DBLP:conf/isca/PatelCZSGMB24} and DistServe \cite{DBLP:conf/osdi/ZhongLCHZL0024} adopt a disaggregated architecture, allocating specialized hardware configurations for each phase. 
Additionally, frameworks like FlexGen \cite{DBLP:conf/icml/0007ZYLRCLRSZ23} and InfiniGen \cite{DBLP:conf/osdi/LeeLSS24} tackle GPU memory constraints through adaptive techniques such as memory compression and intelligent offloading, ensuring scalable performance under resource limitations.
Nevertheless, these systems \gty{fail to}{neither consider the characteristic of pipeline parallelism non} achieve balanced schedule between each batch, leading to significant performance bottleneck. 

% Orca \cite{DBLP:conf/osdi/YuJKKC22}

% vLLM \cite{DBLP:conf/sosp/KwonLZ0ZY0ZS23}
% sglang \cite{DBLP:conf/nips/ZhengYXS0YCKSGB24}

% Splitwise \cite{DBLP:conf/isca/PatelCZSGMB24}
% Distserve \cite{DBLP:conf/osdi/ZhongLCHZL0024}

% flexgen \cite{DBLP:conf/icml/0007ZYLRCLRSZ23}
% infinigen \cite{DBLP:conf/osdi/LeeLSS24}

\textbf{Model Parallelism for LLMs training and serving.}
Model parallelism has become essential for distributed training and serving of LLMs as their size increases. 
Tensor parallelism, which demands frequent communication between devices, is primarily employed in environments with high-bandwidth interconnects. 
Recent advancements \cite{DBLP:conf/asplos/JangdaHLSMMMMS22,DBLP:conf/asplos/WangWSDIHCMMZKG23,DBLP:conf/asplos/ChenLZDSZY24,DBLP:conf/ppopp/DuWJCHCL24} address communication idling by strategically overlapping communication operations with computation. 
For pipeline parallelism, research focuses on solving unbalanced memory consumption \cite{DBLP:conf/asplos/SunCWF0WC24,DBLP:conf/sc/LiuCZ023,DBLP:conf/icml/KimKYC23}, pipeline bubbles \cite{DBLP:conf/sc/LiuCZ023,DBLP:conf/iclr/QiWHL24}, communication optimization \cite{DBLP:conf/ppopp/LinL0WZZ25} and activation checkpointing \cite{DBLP:conf/asplos/SunCWF0WC24,DBLP:conf/ppopp/LiuLTJ25}.
Hybrid strategies combining tensor pipellelism and pipeline parallelism leverage automated search algorithms \cite{DBLP:conf/osdi/ZhengLZZCHWXZXG22,DBLP:conf/eurosys/ZhangD0CW024,DBLP:conf/usenix/UmOKLKKKMJ24} or utilize the heterogeneous characteristics \cite{DBLP:conf/eurosys/ZhangD0CW024,DBLP:conf/usenix/UmOKLKKKMJ24,DBLP:conf/usenix/JiaJWXS0LCLZL022,DBLP:conf/icml/RyabininDDB23}, while frameworks like Megatron-LM \cite{DBLP:conf/sc/NarayananSCLPKV21} demonstrate empirically validated configurations for massive-scale deployment. 
However, those methods focus on optimization during training.
In LLM serving, specialized optimizations target hardware heterogeneity through GPU-aware allocation \cite{DBLP:journals/corr/abs-2408-00741,DBLP:journals/corr/abs-2406-01566} and reduce pipeline bubbles using chunked prefill mechanisms \cite{DBLP:conf/osdi/AgrawalKPMKGTR24}. Nevertheless, pipeline bubbles are not efficiently solved by chunked prefill and become a bottleneck for efficient deploying.

% %%%%%% pipeline parallelism %%%%%%%
% % --- train ---
% Adapipe \cite{DBLP:conf/asplos/SunCWF0WC24} % unbalanced memory consumption
% Hanayo \cite{DBLP:conf/sc/LiuCZ023} %  pipeline bubbles and excessive memory consumption
% Weipipe \cite{DBLP:conf/ppopp/LinL0WZZ25} % communication 
% Mario \cite{DBLP:conf/ppopp/LiuLTJ25} % activation checkpointing
% Zero bubble \cite{DBLP:conf/iclr/QiWHL24} % bubble
% Bpipe \cite{DBLP:conf/icml/KimKYC23} % memory imbalance
% megatron-lm \cite{DBLP:conf/sc/NarayananSCLPKV21} % composed 

% % --- serve ---
% sarathi-serve \cite{DBLP:conf/osdi/AgrawalKPMKGTR24}

% %%%%%% communication %%%%%%%
% CoCoNet \cite{DBLP:conf/asplos/JangdaHLSMMMMS22}
% % --- tensor ---
% \cite{DBLP:conf/asplos/WangWSDIHCMMZKG23}
% Centauri \cite{DBLP:conf/asplos/ChenLZDSZY24}
% Liger \cite{DBLP:conf/ppopp/DuWJCHCL24}

% %%%%%% model parallelism %%%%%%%
% Alpa \cite{DBLP:conf/osdi/ZhengLZZCHWXZXG22} % automate

% %%%%%% heterogeneous %%%%%%%
% % --- train ---
% HAP \cite{DBLP:conf/eurosys/ZhangD0CW024} % automate
% Metis \cite{DBLP:conf/usenix/UmOKLKKKMJ24} % automate
% Whale \cite{DBLP:conf/usenix/JiaJWXS0LCLZL022}
% SWARM Parallelism \cite{DBLP:conf/icml/RyabininDDB23}

% % --- serve ---
% DynamoLLM \cite{DBLP:journals/corr/abs-2408-00741}
% Helix \cite{DBLP:journals/corr/abs-2406-01566}

%% file: chapters/8.conclusion.tex
\section{CONCLUSION}

This paper presents \NAME, a global balanced pipeline parallelism system for distributed LLM serving incorporating with \METHODT~method. \METHODT~dynamically adjusts the token count for prefill or decode stages respectively, ensuring balanced computation across batches and minimizing pipeline bubbles. To enable \METHODT, we design an asynchronous runtime tailored for pipeline parallelism. Experiments with leading LLMs demonstrate that \NAME~delivers 11\% to 398\% higher maximum throughput over state-of-the-art pipeline or tensor parallelism systems, while simultaneously maintaining lower latency. 